\documentclass[poms,final,nonblindrev]{poms1_V1} 

\OneAndAHalfSpacedXI 

\usepackage{algorithm}
\usepackage{algpseudocode}
\usepackage{tikz}

\usepackage{setspace}

\usepackage{bm}
\usepackage{comment}
\usepackage{amsmath}
\usepackage{etoolbox}
\usepackage{float}
\usepackage{booktabs}
\usepackage{graphicx}
\usepackage{multirow}
\usepackage{threeparttable}
\usepackage{caption}
\usepackage[labelfont=sf]{subcaption}
\usepackage{tikz}

\newcommand{\cropimg}[3][0.62]{%
\begin{tikzpicture}[x=\linewidth,y=\linewidth]
    \path[use as bounding box] (0,0) rectangle (1,#1);
    \clip (0,0) rectangle (1,#1);
    \node[anchor=center,inner sep=0pt,yshift=#3] at (0.5,{#1/2})
        {\includegraphics[width=\linewidth]{#2}};
\end{tikzpicture}%
}

\makeatletter
\everydisplay\expandafter{%
  \the\everydisplay
  \abovedisplayskip      6pt plus 2pt minus 2pt\relax
  \belowdisplayskip      6pt plus 2pt minus 2pt\relax
  \abovedisplayshortskip 3pt plus 1pt minus 1pt\relax
  \belowdisplayshortskip 3pt plus 1pt minus 1pt\relax
  \jot=4pt\relax
}
\makeatother

\usepackage[normalem]{ulem}

\definecolor{PLadd}{RGB}{0,100,0}
\definecolor{PLdel}{RGB}{230,120,20}

\usepackage{epsfig}
\usepackage{natbib}
\usepackage{xurl}

\makeatletter
\def\doi#1{\@ifnextchar.{\@gobble}{}}
\makeatother

 \bibpunct[, ]{(}{)}{,}{a}{}{,}%
 \def\bibfont{\small}%

\usepackage{bibunits}
\defaultbibliographystyle{pomsref}
\defaultbibliography{reference}

\usepackage{longtable}
\usepackage{array}
\newcolumntype{L}[1]{>{\raggedright\arraybackslash}p{#1}}

\TheoremsNumberedThrough     
\ECRepeatTheorems

\EquationsNumberedThrough    

\begin{document}


\RUNAUTHOR{Lin et al.}

\RUNTITLE{Service Zone Design for Energy-Constrained Spatial Queueing Systems}

\TITLE{On-Demand Service Zone Design for Energy-Constrained Spatial Queueing Systems}

\ARTICLEAUTHORS{%
\AUTHOR{Peng Lin}
\AFF{Department of Industrial Engineering, Tsinghua University, \EMAIL{lp23@mails.tsinghua.edu.cn}} 
\AUTHOR{Cheng Hua}
\AFF{Antai College of Economics and Management, Shanghai Jiao Tong University, \EMAIL{cheng.hua@sjtu.edu.cn}}
\AUTHOR{Wei Qi}
\AFF{Department of Industrial Engineering, Tsinghua University, \EMAIL{qiw@tsinghua.edu.cn}}
\AUTHOR{Kai Wang}
\AFF{School of Vehicle and Mobility, Tsinghua University, \EMAIL{cwangkai@tsinghua.edu.cn}}
} 

\ABSTRACT{%
Electric service vehicles (ESVs), such as mobile chargers and drone-based service units, are becoming an important operational resource for on-demand service systems. Unlike conventional spatial servers, ESV operations are shaped by battery limits and recharging needs, which affect dispatch feasibility and spatial deployment decisions. We develop an energy-constrained hypercube spatial queueing model that embeds battery-state dynamics into the classical hypercube framework and uses a semi-Markov representation to estimate steady-state performance. We then formulate a joint location--zoning problem for station placement and service zone design. The resulting model is a large-scale mixed-integer nonlinear program, whose set partitioning reformulation has objective coefficients that are not available in closed form. We therefore develop a Branch-Price-and-Evaluation framework for set partitioning problems with externally computable column coefficients: upper-bounding surrogates guide pricing, and iterative exact evaluation updates the coefficients of active columns. Computational results show that explicit energy modeling significantly reduces false service promises and yields more credible planning decisions. They also reveal a load-dependent reversal in zoning: pooling is preferable under light demand, whereas tighter zoning becomes more profitable as demand increases. Over the tested range, profitability is driven more by zoning than by battery improvement, suggesting that managers should get service zone design right before investing in battery upgrades; this caution is reinforced by the counterintuitive finding that larger batteries may delay replenishment and reduce fleet readiness under sparse demand. These findings highlight a broader planning insight for energy-constrained mobile service systems: energy feasibility should be treated not merely as a matter of battery-capacity expansion, but as a design dimension that shapes service-zone configuration.
}%


\newif\ifNoJournalVersion
\NoJournalVersiontrue     

\ifNoJournalVersion
  \JOURNAL{}
  \makeatletter
  \def\theARTICLETOPLEFT{}
  \def\theARTICLETOPRIGHT{}

  \def\theARTICLETOP{\vspace*{-25pt}}

  \RRHFirstLine{}
  \LRHFirstLine{}
  \RRHSecondLine{\bf\theRUNAUTHOR: \it\theRUNTITLE\vspace{1mm}}
  \LRHSecondLine{\bf\theRUNAUTHOR: \it\theRUNTITLE\vspace{1mm}}
  \makeatother
\fi

\maketitle

%
\begin{bibunit}[pomsref]

\section{Introduction}
\begin{figure}[t]
\centering

\begin{subfigure}[t]{0.32\linewidth}
    \centering
    \cropimg[0.62]{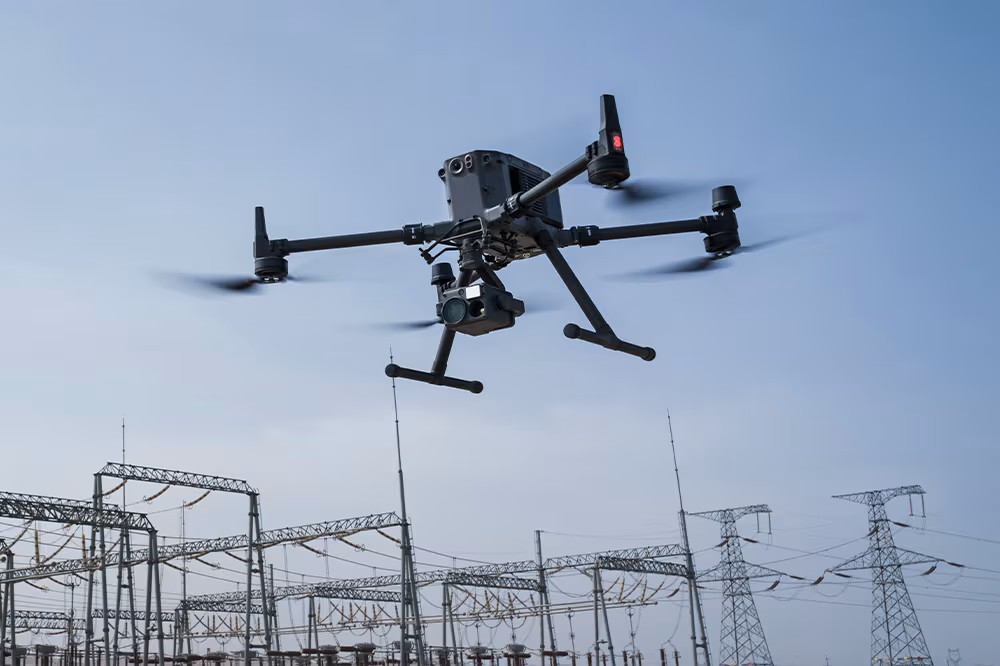}{-2pt}
    \caption{Inspection drone}
    \label{fig:esv_drone}
\end{subfigure}
\hfill
\begin{subfigure}[t]{0.32\linewidth}
    \centering
    \cropimg[0.62]{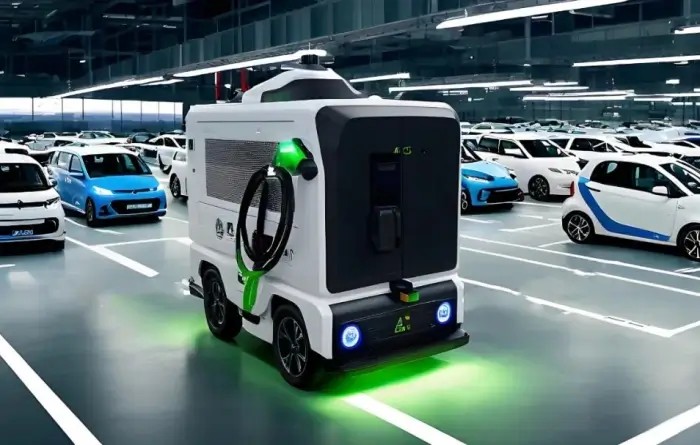}{0pt}
    \caption{Mobile charging pile}
    \label{fig:esv_charge}
\end{subfigure}
\hfill
\begin{subfigure}[t]{0.32\linewidth}
    \centering
    \cropimg[0.62]{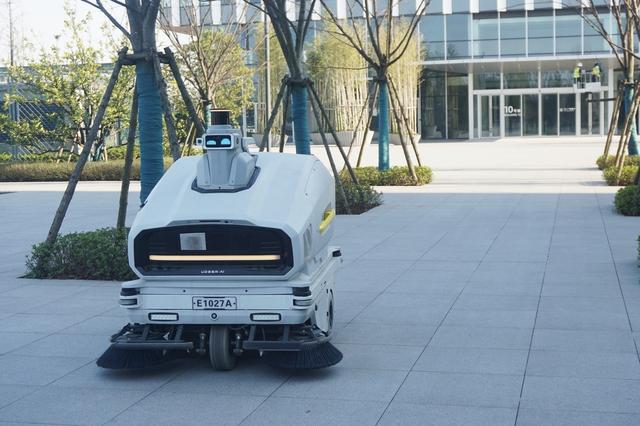}{0pt}
    \caption{Autonomous cleaning robot}
    \label{fig:esv_cleaning}
\end{subfigure}

\vspace{6pt}
\caption{Representative applications of ESVs.}
\label{fig:esv_examples}
\end{figure}

The expansion of on-demand service systems, together with advances in electrification and automation, is reshaping service delivery from fixed-site operations toward mobile and distributed service provision. Electric service vehicles (ESVs) are rapidly emerging as key enablers of this transition. These mobile, battery-powered service units, including mobile charging vans, inspection drones, and even autonomous cleaning robots (see Figure~\ref{fig:esv_examples}), offer unprecedented flexibility in how and where services can be delivered.

For instance, mobile charging vehicles, such as mobile charging vans, are emerging as a promising supplement to fixed electric vehicle (EV) charging infrastructure by providing on-demand charging support, helping mitigate range anxiety, and improving service accessibility in areas with sparse charging coverage \citep{QiuDu2023}. Inspection drones are increasingly deployed to collect visual and sensing data for infrastructure monitoring and maintenance, especially in hard-to-reach or hazardous locations, thereby improving inspection safety, efficiency, and data quality \citep{GrayEtAl2023UAS}. The expanding application of ESVs in logistics, transportation, and public services underscores a critical need for operations models that can both describe and optimize their performance. 

As on-demand energy-based services expand across urban systems, their reliable operations increasingly depend on how mobile service resources are deployed and managed over space and time. In ESV systems, this challenge is amplified by the fact that customer requests may arise at different locations and times, while each vehicle can serve only a limited amount of demand before it must be replenished. Service providers therefore need to understand not only how to plan individual routes or schedules under energy constraints \citep{ChengEtAl2024RobustDroneWeather}, but also how repeated demand arrivals and recurring vehicle recharging jointly affect long-run service performance. This perspective motivates a spatial queueing framework. 

Classical spatial queueing models, however, do not directly capture the energy constraints that are central to ESV operations. Unlike traditional service vehicles, ESVs are constrained by limited battery capacities, which impose a hard limit on their continuous service duration and range \citep{DesaulniersEtAl2016EVRPTW}. An ESV (such as a mobile charger or a drone) can only travel a certain distance or serve a certain number of calls before it must leave its zone to recharge once its energy falls below a critical threshold. During these recharge periods, the vehicle is temporarily out of service, forcing the remaining active servers to cover its area. This coupling between spatial dispatch and energy replenishment creates complex interdependencies: the availability of one server affects others’ utilization in nonlinear ways as the system evolves. From a modeling perspective, energy constraints violate key assumptions of standard queueing theory (e.g., infinite server uptime or exponential service times). Therefore, new models are needed to capture the joint evolution of each server’s location and energy state, as well as their coupled dispatch probabilities, to accurately assess performance metrics in ESV systems. 

In addition to performance modeling, the rise of ESVs raises challenging prescriptive questions in service network design. Operators need to decide how to deploy ESVs across a service region: how to partition the city into service zones for each vehicle or station, how many ESVs to allocate to each zone, and where to locate charging stations or depots. These decisions impact both the spatial coverage of demand and the energy efficiency of operations. Notably, energy limitations mean that an ESV’s effective service radius is restricted; an ESV can only reliably serve requests within a certain distance before needing to return to recharge. If zones are too large or stations are too sparse, vehicles may waste time traveling or may be unable to respond before depleting their battery. On the other hand, placing many stations or overly small zones incurs high setup costs and under-utilizes ESV capacity.  

In this paper, we address both the descriptive and prescriptive challenges above by developing two integrated models and two solution approaches. First, we introduce an energy-constrained hypercube queueing model that descriptively captures the stochastic operations of an ESV system. This model extends the classical hypercube framework \citep{Larson1975ApproxEMS} by incorporating battery state dynamics into the server dispatch process. We model each ESV as a server in a semi-Markov process that tracks both its spatial state (idle, busy serving a call) and its energy state (remaining battery level) over time. In essence, when an ESV is dispatched to a service call, its battery is depleted according to the task’s service-energy requirement; when its battery falls below a threshold, it is recalled to its dedicated base for replenishment, during which it is unavailable for calls. We represent these coupled location-energy state transitions via an augmented state-space and derive the steady-state distribution of the system using a fixed-point approximation method called Energy-Hypercube Iteration. This allows us to compute key performance metrics for a given deployment of ESVs, for example, service fulfillment, server utilization, and long-run profit rate, while explicitly accounting for energy constraints.

Second, we formulate a joint optimization model for on-demand service zone design that builds on the above descriptive analysis. The model simultaneously determines the optimal locations of ESV stations and the partitioning of the service region into zones assigned to each station. The constraints enforce each demand node to be assigned to exactly one zone, ensure that each zone is served by a pool of ESVs, and impose contiguity on the resulting service zones. This optimization problem can be formulated as a mixed-integer nonlinear program (MINLP), because its objective depends on steady-state performance metrics from the queueing model, and an explicit embedding of these metrics introduces nonlinear constraints. The nonlinearity is inherited not only from the hypercube performance model itself, but also from its coupling with energy dynamics, through multiplicative dispatch terms, ratio-based energy-driven transition probabilities, normalized bilinear occupancy ratios, and implicit rational inverse-Erlang load equations. The resulting formulation is therefore a nonlinear combinatorial location--zoning design problem, in which binary station-location and zone-assignment decisions are coupled with embedded steady-state queueing relations. Evaluating any candidate design thus requires solving the underlying performance system, rendering exact optimization computationally demanding.

To solve this difficult optimization problem, we propose a novel Branch-Price-and-Evaluation framework. This approach is inspired by the branch-and-price method but augmented with a custom column evaluation mechanism to handle the nonlinear aspects of our model. In our algorithm, we decompose the zone design problem by treating each feasible zone as a column in a column-generation framework. The inner loop of our method performs column generation. It solves a relaxed master problem to identify promising service zones (columns) to add, similar to standard branch-and-price techniques. The outer loop then performs column evaluation, wherein we recompute the objective coefficients of columns under the nonlinear queueing model and update the master problem accordingly. By iterating between solving a linear approximation and updating it with accurate performance metrics, and branching on fractional assignments as needed, the Branch-Price-and-Evaluation algorithm converges to an optimal integer solution of the original MINLP. This two-level approach effectively delegates the challenging nonlinear calculations to a specialized evaluation step while leveraging efficient linear solvers for the combinatorial structure. To our knowledge, this is the first integration of a column-generation pricing method with a dynamic evaluation feedback loop within a branch-and-bound tree.

\textbf{Contributions.} This work provides a framework for analyzing and optimizing energy-constrained service systems, with contributions to both methodology and application, as outlined below: 
\begin{itemize}
    \item \textit{Formalizing Energy-Constrained Mobile Service Systems}. We establish ESV-based on-demand services (e.g., drone fleets, mobile charging units) as a distinct class of spatial queueing systems with endogenous energy constraints. We highlight how battery limitations introduce intricate couplings between dispatching and refueling that fundamentally differentiate ESV operations from traditional vehicle-based services. 
    \item \textit{Energy-Constrained Hypercube Queueing Model}. We extend Larson’s seminal hypercube queueing model \citep{Larson1975ApproxEMS} to analyze ESV system performance by incorporating stochastic energy depletion and recharge. The model novelly embeds a semi-Markov process to capture each server’s joint location-energy state transitions and applies a fixed-point iterative scheme to compute steady-state metrics. This is, to the best of our knowledge, the first analytical model to merge spatial queueing with battery dynamics. It enables calculation of metrics such as steady-state service coverage probability and system profit rate, providing an analytical tool for evaluating ESV deployments under various configurations and demand patterns.
    \item \textit{Branch-Price-and-Evaluation Optimization Framework}. We propose a new optimization methodology to prescriptively design ESV service zones, formulated as a MINLP. To solve it, we introduce the Branch-Price-and-Evaluation algorithm, which combines branch-and-price with a two-level column generation and evaluation process. This framework iteratively generates candidate service zones columns and evaluates their performance contribution by embedding the queueing model within the optimization loop. The algorithm can handle the large-scale, nonlinear nature of the problem and guarantees convergence to optimality. This is the first application of such a technique in the context of spatial queueing design.
    \item \textit{Insights for ESV Network Design}. We provide extensive numerical experiments and three real-data-based cases to demonstrate the value of the proposed framework. The experiments yield three managerial insights. First, explicitly modeling energy is critical for credible planning, as energy-agnostic models can overstate service capacity and lead to infeasible deployment decisions. Second, the value of zoning is load-dependent: tighter zoning is more profitable under heavier demand by improving travel efficiency and local service focus. Third, zoning and battery expansion interact in a demand-dependent manner. Although they act as partial substitutes, zoning has a larger profit impact, suggesting that managers should align zoning with the demand regime before investing in additional battery capacity. Counterintuitively, expanding battery capacity is not uniformly beneficial and can reduce fleet readiness under sparse demand, whereas additional capacity is more readily converted into completed services under heavier load.
\end{itemize}

The paper is organized as follows. Section 2 discusses relevant results from the literature. Section 3 presents our energy-constrained hypercube queueing system. Section 4 presents the joint location--zoning model and the Branch-Price-and-Evaluation framework. Section 5 reports computational results and managerial insights. Section 6 concludes with directions for future research. A summary of notation and all technical proofs are provided in the appendices. 

\section{Literature Review}
\label{sec:literature-review}

\subsection{Descriptive Analytics for Energy-Constrained Mobile Service Systems}
\label{subsec:descriptive-analytics}

Energy-constrained mobile service operations arise in applications such as mobile EV charging \citep{Yan2024MobileChargingCrowd}, drone inspection \citep{XiaWangWang2019DroneScheduling}, and autonomous robotic-service fleets \citep{ZouEtAl2018RMFSBattery}. 
These settings share a common operational feature: spatially distributed tasks are served by electric mobile units whose availability depends jointly on location and energy dynamics.
Because service consumes energy and recovery restores future availability, energy is not merely a resource constraint but an endogenous state variable that affects both current service feasibility and future fleet capacity. 
Existing application models often recognize energy limitations, but typically do not represent energy as a stochastic state of distinguishable mobile servers evolving through repeated service and recovery cycles.

Much of the related literature uses routing or scheduling formulations, which are appropriate for planning over a finite task set or service horizon \citep{QiuDu2023,ChengEtAl2024RobustDroneWeather}. 
Our objective is instead to evaluate the long-run performance, such as steady-state workloads, availability, and dispatch frequencies. 
This shifts the modeling focus from finite-horizon route optimization to spatial queueing-based analytics.

Queueing models are therefore a natural framework for analyzing congestion and availability in electric mobile service systems. 
However, most existing queueing work in this area is facility-centric, focusing on charging stations, depots, or other fixed service facilities rather than on the mobile servers. For example, \citet{HeEtAl2021EVChargingFleet} model an EV-sharing fleet as a multiclass open queueing network in which vehicles flow among demand zones, roads, and charging sites, with classes defined by battery energy levels. 
\citet{QiEtAl2023BatterySwapping} model urban battery swapping as a repairable-inventory queueing system that captures the stochastic circulation of depleted and charged batteries between swapping and charging stations. 
However, their state descriptions are aggregate at the zone, facility, or inventory-network level, whereas on-demand ESV operations require tracking the state of each mobile server because dispatch feasibility depends on both server identity and residual energy.

The hypercube queueing model provides the closest methodological foundation for our system. Introduced by \citet{Larson1974Hypercube,Larson1975ApproxEMS}, the hypercube model tracks the joint busy-idle state of distinguishable servers under spatially distributed demand and preference-based dispatch. Recent applications have used this framework for urban service design, such as police-zone design \citep{ZhuWangXie2022Ambulance}. In particular, \citet{HuaSwersey2022FireMedics} develop a three-state spatial queueing model for cross-trained fire-medics. Our setting differs in that the additional state is an evolving energy stock: after completing a request, an ESV may be fully available, feasible only for a subset of requests, or forced into recovery before future dispatch. This paper extends the hypercube framework to incorporate residual-energy-dependent service feasibility and recovery-induced unavailability into steady-state spatial dispatch analysis.

\subsection{Prescriptive Analytics for Service-Zone Design}
\label{subsec:prescriptive-analytics}

The prescriptive side of our problem concerns the design of service zones and facility locations that induce high-profit ESV operations. Related work falls broadly into two streams: service region design and districting, which we refer to as zoning in our setting. In service-region design, the central decisions typically concern which demand areas to serve and how to locate or provision facilities to satisfy coverage and service-level requirements \citep{Larson1974Hypercube,BaronEtAl2008FacilityLocation,BaronEtAl2009Feasibility,HeEtAl2017EVSharingRegion}. 
In contrast, districting partitions a prescribed region into mutually exclusive zones and exactly assigns all demand units to one of them, often subject to contiguity, compactness, and balance requirements \citep{HessEtAl1965Redistricting,shirabe2009districting,RiosMercadoFernandez2009GRASP}. Logistics and delivery applications extend this perspective by evaluating zones through workload and service reliability \citep{HauglandHoLaporte2007DeliveryDistricts,BanerjeeEtAl2022FleetSizing,CarlssonDelage2013RobustPartitioning}. Our study is closer to this districting stream: it partitions all demand nodes into service zones, assigns each opened base to one zone, and restricts dispatching within zones, with the objective of selecting zones that maximize system-wide long-run profit. 

This partitioning structure naturally leads to a set-partitioning formulation in which each feasible service zone is represented as a column. Column generation is a standard approach for solving large-scale set-partitioning relaxations, and branch-and-price embeds column generation within branch-and-bound to obtain integer solutions \citep{BarnhartEtAl1998BranchPrice,LuebbeckeDesrosiers2005ColGen}. 
This column-based paradigm has also been applied directly to districting problems, with columns encoding feasible districts or district-level assignment plans \citep{MehrotraEtAl1998PoliticalDistricting,ZhenEtAl2023TerritorialDesign}.
In these settings, however, the contribution of a generated column is an explicit cost or performance coefficient that can be computed within the pricing problem.

A closer methodological connection arises when column coefficients are difficult to evaluate. 
A common strategy in this stream is to transfer nonlinearities from the master problem to the pricing problem, and then exploit the structure of the resulting pricing problem, such as submodularity \citep{TeoShu2004WarehouseRetailer}, aggregate nonlinear facility-cost structure \citep{NiEtAl2021FacilityLocation}, or dynamic programming recursions \citep{JacquillatEtAl2025ContagionBAP}. 
The closest work to ours is \citet{ZhangEtAl2023ScenarioReduction}, who study large-scale stochastic programs with hard-to-estimate objective parameters and develop a column-evaluation-and-generation approach that alternates between generating columns using relaxation-based approximations and evaluating the true objective coefficients of selected columns. 
Our setting differs from this literature in that the contribution of a column is an implicit quantity obtained through external evaluation rather than a structured objective term that can be optimized directly in the pricing problem.
We therefore develop a Branch-Price-and-Evaluation framework in which tractable surrogate coefficients guide column generation, and exact energy-constrained hypercube evaluation updates the coefficients of generated service-zone columns.

\section{The Energy-Constrained Hypercube Queueing System}

\subsection{Problem Definition}

\begin{figure}[t]
    \centering
    \newcommand{\figH}{6.7cm}

    \begin{subfigure}[b]{0.44\textwidth}
        \centering
        \includegraphics[
            height=\figH
        ]{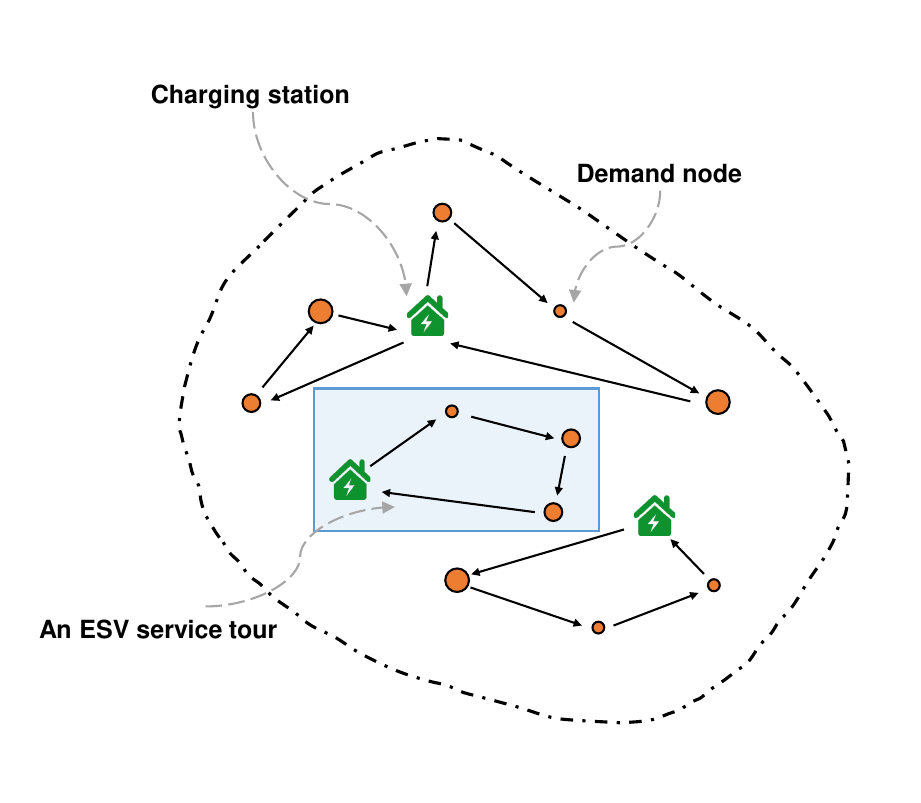}
        \caption{Spatial structure of ESV service system.}
        \label{fig:spatial_structure}
    \end{subfigure}
    \hfill
    \begin{subfigure}[b]{0.55\textwidth}
        \centering
        \includegraphics[
            height=\figH
        ]{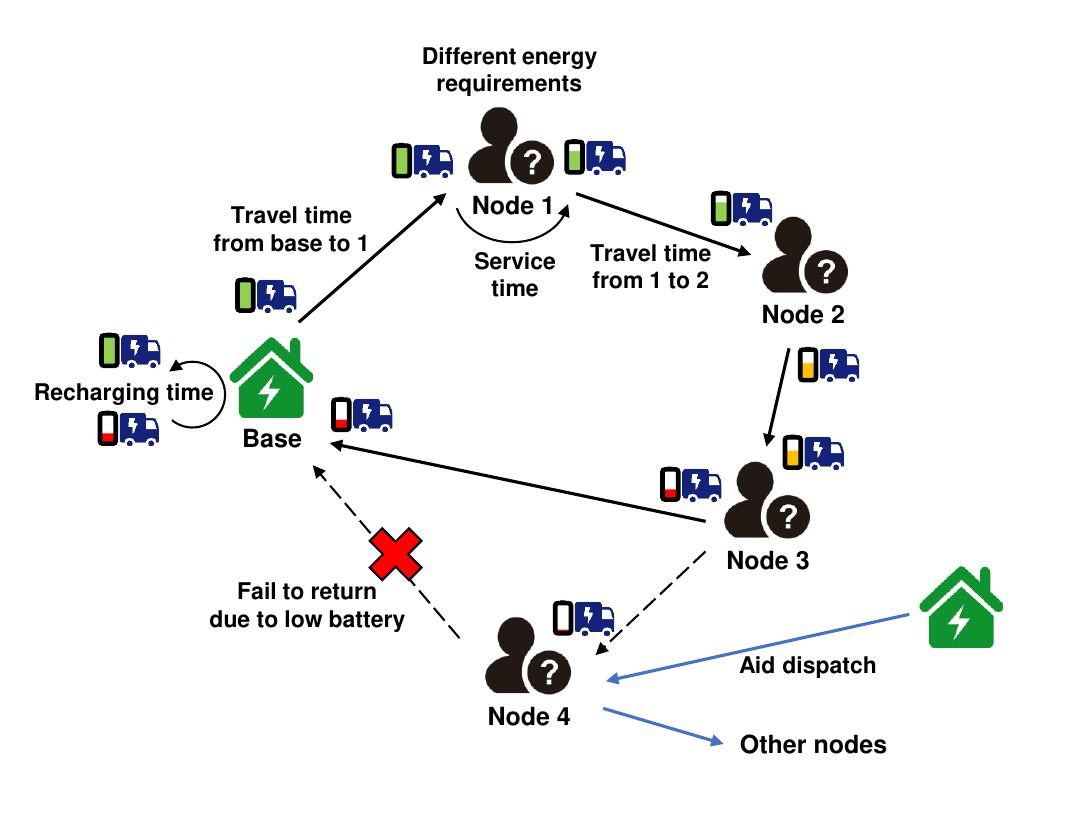}
        \caption{Single-unit operation mechanism.}
        \label{fig:single_unit_mechanism}
    \end{subfigure}

    \caption{The spatial structure and single-unit operation mechanism of ESV service system. In panel (a), the sizes of the circular demand nodes represent different energy requirements.}
    \label{fig:system_overall_illustration}
\end{figure}

We consider an on-demand ESV service system in which customer requests differ in their energy requirements, as shown in Figure~\ref{fig:system_overall_illustration}\subref{fig:spatial_structure}. Demand is spatially aggregated into $J$ demand nodes, indexed by $j=1,\dots,J$, and is served by $N$ ESV units, indexed by $i=1,\dots,N$. Each ESV unit $i$ starts a service tour from a dedicated charging station (base) $b_i$, serves several demand nodes, returns to its base for recharging, and waits for next demand. We focus on ESV applications in which energy consumed in travel is relatively small and predictable compared with the energy required for on-site service. Accordingly, before mapping the physical battery capacity to the discretized state space, we deduct a calibrated allowance for travel and return movements. The remaining effective capacity is then used to define the energy levels available for customer service. The battery state of each unit is then discretized into $E+1$ levels, $\{0,1,\dots,E\}$, where $E$ denotes the maximum effective energy capacity and $0$ denotes depletion of this dispatchable energy budget. In operation, however, units are not allowed to deplete their batteries completely, so level $0$ is effectively never reached. 

A demand arriving at node $j$ and requiring $e$ units of energy, where $e=1,\dots,E-1$, is referred to as a type-$(j,e)$ demand and arrives according to a Poisson process with rate $\lambda_{je}$. The total arrival rate at node $j$ is therefore $\lambda_j=\sum_{e=1}^{E-1}\lambda_{je}$. Each demand node $j$ is associated with a predetermined preference list over units, and $\gamma_{jk}$ denotes the index of the $k$th preferred unit for node $j$. When a call arrives at node $j$, the request is offered sequentially according to this priority order until a unit that is free and has sufficient battery accepts the dispatch; otherwise, the demand is lost.

We do not assume exponential service times. Instead, each service tour is decomposed into a travel component, an on-site service component, and, when necessary, a return-and-recharge component. In this study, the duration of each component is treated as deterministic conditional on the realized service tour; however, the overall service-tour duration remains random and generally non-exponential because the realized demand type and the resulting tour are random. Let $\tau^\mathrm{t}_{jj'}$ denote the travel time from node $j$ to node $j'$, and let $\tau^\mathrm{s}_e$ denote the on-site service time for a demand requiring $e$ units of energy. For unit $i$, let $\tau^\mathrm{rt}_{ij}$ denote the travel time from node $j$ back to its base $b_i$, and let $\tau^\mathrm{rc}_e$ denote the recharging time from battery level $e$ to full charge $E$. The superscripts $\mathrm{t}$, $\mathrm{s}$, $\mathrm{rt}$, and $\mathrm{rc}$ stand for travel, service, return, and recharge respectively. Because the semi-Markov analysis in Section~\ref{sec:semi-Markov} is carried out for a fixed unit $i$, we suppress the unit index and write $\tau^\mathrm{rt}_j$ for $\tau^\mathrm{rt}_{ij}$ whenever no confusion arises.

We next describe a typical service tour of a unit as Fig~\ref{fig:system_overall_illustration}\subref{fig:single_unit_mechanism} illustrates. Let $(e,j)$ denote the state in which the unit is located at node $j$ with battery level $e$. In particular, each service tour of unit $i$ starts from its dedicated base $b_i$ in state $(E,b_i)$. More generally, suppose that a unit is in state $(e,j)$ when it is dispatched to a type-$(j',e')$ demand. The unit immediately starts traveling to node $j'$, arriving after $\tau^\mathrm{t}_{jj'}$. The unit then provides on-site service for $\tau^\mathrm{s}_{e'}$, after which its battery level decreases from $e$ to $e-e'$. Accordingly, upon service completion, the unit is in state $(e-e',j')$. The transition from $(e,j)$ to $(e-e',j')$ takes $\tau^\mathrm{t}_{jj'}+\tau^\mathrm{s}_{e'}$. If the remaining battery level satisfies $e-e'\ge \theta$, the unit stays at node $j'$ and waits for the next dispatch. Otherwise, it is recalled to its base $b_i$, returns in time $\tau^\mathrm{rt}_{j'}$, and is then recharged to level $E$, which requires an additional time $\tau^\mathrm{rc}_{e-e'}$. After recharging, the unit returns to state $(E,b_i)$.

We distinguish between a unit being \emph{idle} and \emph{available}. A unit is idle if it is neither serving a request nor undergoing replenishment. An idle unit is said to be available for a type-$(j,e)$ demand if its current battery level is sufficient to complete that service; otherwise, it is unavailable for that demand type.

This setting poses three fundamental challenges for spatial queueing analysis. First, dispatch feasibility is both state-dependent and demand-dependent: whether a unit is available for an arriving request depends not only on whether it is idle, but also on its location, residual battery, and the energy requirement of the realized demand. Second, idle units are no longer homogeneous: two units that are both idle may still differ in the set of demands for which they are available, because their residual battery levels may differ. Third, service completion does not return a unit to a common post-service state: depending on its remaining battery, the unit may either stay in the field and become available for future dispatches or return to its base for recharging before becoming available again. If these features are ignored, the model will mischaracterize the set of requests the system can feasibly serve and thereby overstate system performance. Departing from the classical hypercube model, we develop an energy-constrained hypercube queueing system that captures these features by augmenting the unit state with battery information and explicitly modeling battery depletion and recharging. Following \cite{HuaSwersey2022FireMedics}, we focus on the stationary performance of the resulting system.

\subsection{Stationary Approximation}

Consider a type-$(j,e)$ demand arrival, let $p_{ije}$ denote the probability that unit $i$ is dispatched to serve it. 
This dispatch probability depends on both the utilization of each unit and the distribution of its battery level when idle. Let $\rho_i$ denote the utilization of unit $i$, i.e., the long-run probability that unit $i$ is busy, either serving a demand or recharging. For an idle unit $i$, let $p_i^e$ denote the probability that its battery level is $e$. Define
$
p_i^{>e}=p_i^{e+1}+\cdots+p_i^E
$
and 
$
p_i^{\le e}=1-p_i^{>e}.
$
It is then convenient to define
$
\rho_i^e=\rho_i+(1-\rho_i)p_i^{\le e},
$
which is the probability that unit $i$ is unavailable to a type-$(j,e)$ demand, either because it is busy or because it is idle but energy-insufficient. We further define
$
\bar \rho^e=\frac{1}{N}\sum_{i=1}^N \rho_i^e,
$
the average probability that a unit is unavailable to a type-$(j,e)$ demand.
With this notation, the dispatch probability is approximated by
\begin{equation}
\begin{split}
p_{ije}
&\approx Q(N,\bar \rho^e,k-1)\prod_{l=1}^{k-1}\rho_{\gamma_{jl}}^{ e}\cdot (1-\rho_i^{e}),
\end{split}
\label{eq:disp_prob}
\end{equation}
where $i=\gamma_{jk}$, that is, unit $i$ is the $k$th preferred unit for node $j$. 

The logic of \eqref{eq:disp_prob} follows from the dispatch rule. For unit $i$ to be dispatched to a type-$(j,e)$ demand, every unit ranked ahead of $i$ in node $j$'s preference list must be unavailable to that demand, whereas unit $i$ itself must be idle and have sufficient battery. Following the classical hypercube approximation of \cite{Larson1975ApproxEMS}, a correction factor is introduced to partially account for dependence among unit states. Specifically,
\begin{equation}
Q(N,\bar\rho,k)=\sum_{l=k}^{N-1}\frac{\binom{l}{k}}{\binom{N}{k}}\frac{N-l}{N-k}\frac{P(l)}{\bar\rho^k(1-\bar\rho)},
\label{eq:Q}
\end{equation}
where $P(l)$ denotes the probability that exactly $l$ units are unavailable, and $\bar\rho$ is the average unavailability probability. In our setting, unavailability is defined relative to a type-$(j,e)$ demand, so the correction is applied with the demand-specific quantity $\bar\rho=\bar\rho^e$, and the distribution $P(l)$ is specialized to $P^e(l)$. Both $\bar\rho^e$ and $P^e(l)$ are discussed further in Section~\ref{sec:inverse_erlang}. In particular, $Q(N,\bar\rho,0)=1$. 

To evaluate \eqref{eq:disp_prob}, we develop, for each unit, a semi-Markov process to determine $\rho_i$ and $p_i^e$ in Section~\ref{sec:semi-Markov}, and propose an inverse Erlang method to derive the correction factor $Q$ in Section~\ref{sec:inverse_erlang}. 

\subsubsection{Semi-Markov Process for Unit-State Transitions.}\label{sec:semi-Markov}

Fix a unit $i$ and observe it immediately after each service completion or recharging completion. The resulting state is denoted by $(e,j)$, where $e$ is the battery level and $j$ is the unit's location. Although demands arrive exogenously to demand nodes at rates $\lambda_{je}$, from the perspective of a unit these arrivals are transformed by the dispatch rule into offer rates, which govern the transition of its $(e,j)$ state.

\begin{figure}[!ht]
    \centering
    \includegraphics[scale=0.50]{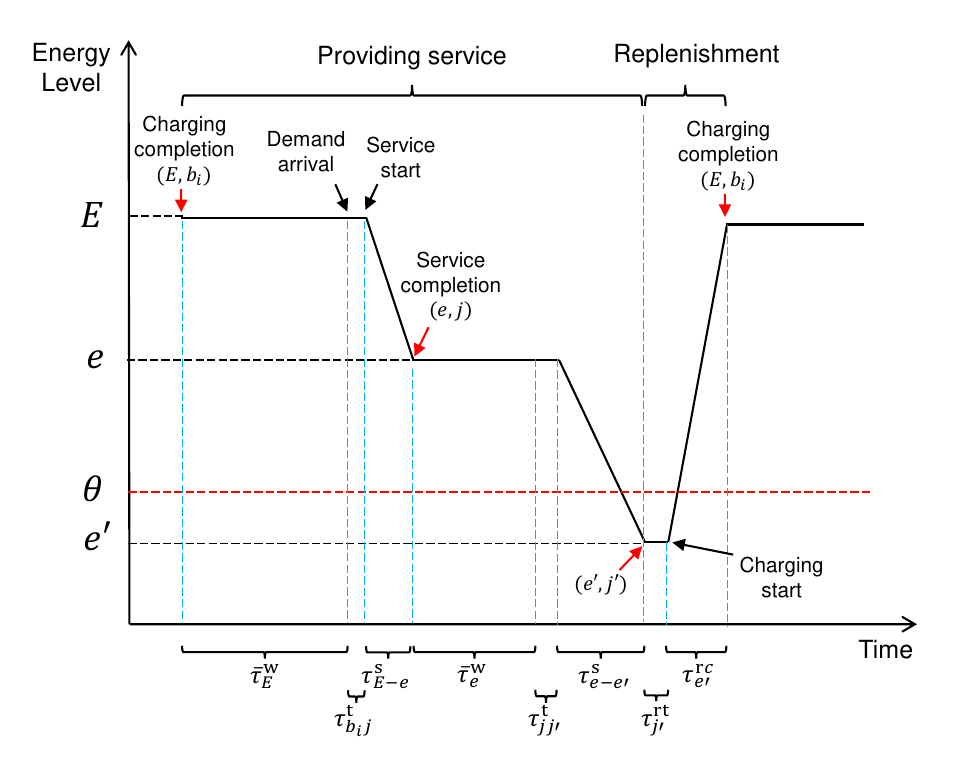}
    \caption{Illustration of the post-completion state evolution of a fixed unit. The state $(e,j)$ records the battery level and location immediately after a service completion or a recharging completion, and the dashed vertical lines mark these epochs. When $e>\theta$, the next offered demand of type-$(j',e-e')$ moves the state from $(e,j)$ to $(e',j')$; when $e\le\theta$, the unit returns to base, recharges to level $E$, and re-enters at $(E,b_i)$. The labels on the time axis indicate the corresponding components of the sojourn time.}
    \label{fig:state_transition}
\end{figure}

To characterize the competition faced by an idle unit, let $O_{ije}$ denote the rate at which type-$(j,e)$ demands are offered to unit $i$, regardless of whether unit $i$ has sufficient battery to accept. If $i=\gamma_{jk}$, then
\begin{equation}
    O_{ije} = \lambda_{je}Q(N,\bar\rho^e,k-1)\prod_{l=1}^{k-1}\rho_{\gamma_{jl}}^{e}.
    \label{eq:offer_rate}
\end{equation}
We further define
$O_{ie}=\sum_j O_{ije},$
to be the total offer rate to unit $i$ for demands requiring $e$ levels of energy.

Let $\mathbf{P}^i$ denote the state-to-state transition matrix governing the next post-completion state of unit $i$. Because the analysis in this subsection is carried out for a fixed unit $i$, we omit the index $i$ from the transition probabilities below. The nonzero entries of $\mathbf{P}^i$ are
\begin{equation}
\begin{split}
    \mathbf{P}_{(e,j)(e',j')} &= \frac{O_{ij'(e-e')}}{\sum_{\epsilon=1}^{e-1} O_{i\epsilon}}, \quad e>\theta,\ e>e'\geq 1,\\
    \mathbf{P}_{(e,j)(E,b_i)} &= 1, \quad e\leq\theta,
    \label{eq:Markov_transition}
\end{split}
\end{equation}
and all remaining elements are zero. Figure~\ref{fig:state_transition} visualizes the two cases in \eqref{eq:Markov_transition}. Under the mild assumption that each demand node has a complete preference list over all units and that $\lambda_{j1}>0$ for all demand nodes $j$, we have $O_{i1}>0$ and hence $\sum_{\epsilon=1}^{e-1} O_{i\epsilon}>0$ whenever $e>\theta$. Thus, \eqref{eq:Markov_transition} is well defined.

Figure~\ref{fig:state_transition} also indicates the waiting, travel-service, and travel-recharging components entering the sojourn times below. Let $\bar \tau_e^\mathrm{w}$ denote the expected waiting time until the next offer arrives when the unit is in a dispatchable state with battery level $e>\theta$. Since the total offer rate out of such a state is $\sum_{\epsilon=1}^{e-1}O_{i\epsilon}$, we have
\begin{equation}
\begin{split}
    \bar \tau_e^\mathrm{w} &= \frac{1}{\sum_{\epsilon=1}^{e-1} O_{i\epsilon}}, \quad e=\theta+1,\dots,E,\\
    \bar \tau_e^\mathrm{w} &= 0, \quad e=1,\dots,\theta.
    \label{eq:sojourn_time}
\end{split}
\end{equation}
Accordingly, the sojourn time associated with a feasible transition is
\begin{equation*}
    h_{(e,j)(e',j')}
    :=
    \begin{cases}
        \bar \tau_e^\mathrm{w} + \tau^\mathrm{t}_{jj'} + \tau^\mathrm{s}_{e-e'}, & e>\theta,\ e>e'\geq 1,\\[4pt]
        \tau^\mathrm{rt}_j + \tau^\mathrm{rc}_e, & e\leq\theta,\ (e',j')=(E,b_i).
    \end{cases}
\end{equation*}
For all state pairs $((e,j),(e',j'))$ such that $\mathbf{P}_{(e,j)(e',j')} = 0$, the value of $h_{(e,j)(e',j')}$ may be defined arbitrarily.

With the transition matrix $\mathbf{P}^i$ and the sojourn times $\{h_{(e,j)(e',j')}\}$ specified, we can now construct a continuous-time process describing the post-completion evolution of unit $i$. Let $\{X_n^i\}_{n\ge 0}$ denote the sequence of post-completion states of unit $i$, and write $X_n^i=(E_n^i,J_n^i)$ for its state at the $n$th post-completion epoch. By construction, $\{X_n^i\}_{n\ge 0}$ is a Markov chain with transition matrix $\mathbf{P}_i$. Let $\{T_n^i\}_{n\ge 0}$ be the corresponding jump times, defined recursively by
$
T_0^i = 0,\; T_{n+1}^i = T_n^i + S_{n+1}^i,\;
S_{n+1}^i := h_{X_n^i X_{n+1}^i},\; n\ge 0.
$
Based on $\{X_n^i\}$ and $\{T_n^i\}$, define the continuous-time process $\{X^i(t)\}_{t\ge 0}$ by
$
X^i(t)=X_n^i,\; t\in [T_n^i,T_{n+1}^i),\; n=0,1,2,\dots.
$
Then $\{X^i(t)\}_{t\ge 0}$ is piecewise constant and changes state only at the jump times. The next proposition summarizes the resulting semi-Markov structure.

\begin{proposition}
The process $\{X^i(t)\}_{t\ge 0}$ is a semi-Markov process with embedded chain $\mathbf P^i$ and semi-Markov kernel
\[
Q^i_{(e,j)(e',j')}(t)
:=
\mathbb{P}\!\left(
    X_{n+1}^i = (e',j'),
    \; T_{n+1}^i - T_n^i \le t
    \,\big|\,
    X_n^i = (e,j)
\right)
=
\mathbf{1}_{\{t \ge h_{(e,j)(e',j')}\}}\,
\mathbf{P}_{(e,j)(e',j')}.
\]
\label{prop:semi_markov}
\end{proposition}
We next use this semi-Markov representation to derive the utilization $\rho_i$. Recall that $\rho_i$ is the long-run fraction of time that unit $i$ is either serving demand or recharging:
\begin{equation}
    \rho_i=\frac{\bar t^\mathrm{s}+\bar t^\mathrm{r}}{\bar t^\mathrm{w}+\bar t^\mathrm{s}+\bar t^\mathrm{r}},
    \label{eq:rho_time_limit}
\end{equation}
where $\bar t^\mathrm{s}$, $\bar t^\mathrm{r}$, and $\bar t^\mathrm{w}$ denote the long-run total travel-service time, travel-recharging time, and waiting time, respectively.
Let $\pi_{(e,j)}$ be the stationary distribution of the embedded chain. It satisfies
\begin{equation}
\begin{split}
    &\bm \pi = \bm \pi \mathbf{P}, \\
    &\bm \pi \mathbf{1} = 1.
    \label{eq:stationary_distribution}
\end{split}
\end{equation}
Together with the time components, we obtain
\begin{equation}
\begin{split}
    \bar t^\mathrm{s} &= \sum\limits_{e>e',\,e>\theta} \sum_{j,j'} \pi_{(e,j)}\mathbf{P}_{(e,j),(e',j')}(\tau^\mathrm{t}_{jj'}+\tau^\mathrm{s}_{e-e'}),\\
    \bar t^\mathrm{r} &= \sum\limits_{e\leq\theta} \sum_j \pi_{(e,j)}(\tau^\mathrm{rt}_j+\tau^\mathrm{rc}_e), \\
    \bar t^\mathrm{w} &= \sum\limits_{e>\theta} \sum_j \pi_{(e,j)}\bar \tau_e^\mathrm{w}.
    \label{eq:time_sum}
\end{split}
\end{equation}
Combining \eqref{eq:rho_time_limit} and \eqref{eq:time_sum} yields $\rho_i$.

The battery-level probabilities $\{p_i^e\}$ are then obtained from the long-run occupancy:
\begin{equation}
\begin{split}
    p_i^e &= \frac{\sum_{j}\pi_{(e,j)}\cdot \bar \tau_e^\mathrm{w}}{\sum_{e'>\theta}\sum_{j} \pi_{(e',j)}\cdot \bar \tau_{e'}^\mathrm{w}}, \quad e=\theta+1,\dots,E, \\
    p_i^e &= 0, \quad e=1,\dots,\theta.
    \label{eq:pie}
\end{split}
\end{equation}

At this point, the only quantity in \eqref{eq:disp_prob} that remains to be determined is $Q(N,\bar\rho^e,k-1)$, which requires the occupancy distribution $\{P^e(k)\}$ of unavailable servers.

\subsubsection{Inverse Erlang Method.}\label{sec:inverse_erlang}

We now develop an inverse Erlang method to obtain a tractable expression for the correction factor $Q(N,\bar\rho^e,k-1)$ in \eqref{eq:disp_prob}. For each energy level $e$, we approximate the unavailability-count distribution by an auxiliary $M/G/N/N$ loss system and recover its offered load via the inverse Erlang method. In this auxiliary system, a unit is treated as busy if it is unavailable to a type-$(j,e)$ demand. Let $\varphi^e$ denote the offered load of this approximate system. Its carried load is
\begin{equation}
    C^e=\sum_{i=1}^N \rho_i^{e} = \varphi^e\bigl(1-P^e(N)\bigr),
    \label{eq:loads}
\end{equation}
where $P^e(N)$ is the blocking probability of the corresponding $M/G/N/N$ loss system. The steady-state probability that exactly $k$ units are unavailable in this approximate system is
\begin{equation}
    P^e(k)=\frac{(\varphi^e)^k/k!}{\sum_{i=0}^N (\varphi^e)^i/i!}.
    \label{eq:P_k}
\end{equation}
Given $\{\rho_i^{e}\}_{i=1}^N$, the carried load $C^e$ is known. We then compute the unique solution $\varphi^e$ to \eqref{eq:loads} by Newton's method (following \citealp{MaHuaSwersey2026ThreeStateEMS}), and obtain $P^e(k)$ from \eqref{eq:P_k}. Moreover, the average unavailability probability across units is
\begin{equation}
    \bar\rho^e=\frac{1}{N}\sum_{k=0}^N k\,P^e(k)=\frac{C^e}{N}.
    \label{eq:r^e}
\end{equation}
Substituting $P^e(k)$ and $\bar\rho^e$ into the correction factor yields
\[
Q(N,\bar\rho^e,k) = \sum_{l=k}^{N-1} \frac{\binom{l}{k}}{\binom{N}{k}}
\frac{N-l}{N-k}\frac{P^e(l)}{(\bar\rho^e)^k(1-\bar\rho^e)},
\]
with $Q(N,\bar\rho^e,0)=1$.

\subsubsection{Energy-Hypercube Iteration.}\label{sec:iter_alg}

\begin{algorithm}[!htbp]
\caption{Energy-Hypercube Iteration}\label{algo:iterative_fixed_point}
{\footnotesize
\begin{algorithmic}[1]
\State \textbf{Input:} $\lambda_{je},\tau^\mathrm{t}_{jj'}, \tau^\mathrm{s}_e, \tau^\mathrm{rc}_e, \forall i\in[N], j\in[J], e\in[E-1]$. 
\State \textbf{Initialize:} Randomize $p_{ije}(0)\in(0,1), \forall i\in[N], j\in[J], e\in[E-1]$, satisfying $\sum_i p_{ije}(0)=1$. Let $O_{ije}(0)=\lambda_{je}p_{ije}(0)$.
\While {$\max_{i,j,e} |O_{ije}(t)-O_{ije}(t-1)| > \epsilon$}
\State Update $\mathbf{P}(t)$ for each $i\in [N]$ from \eqref{eq:Markov_transition}.
\State Update $\rho_i(t), p_i^e(t)$ for each $i\in [N], e\in [E]$ from \eqref{eq:rho_time_limit}--\eqref{eq:pie}. 
\State Update $Q(N, \bar\rho^e,k)(t), \forall e\in[E]$ from \eqref{eq:loads}--\eqref{eq:r^e} by inverse Erlang method. 
\State Update $O_{ije}(t), \forall i\in[N], j\in[J], e\in[E-1]$ from \eqref{eq:offer_rate}.
\EndWhile
\State \textbf{Output:} $p_{ije}, \forall i\in[N], j\in[J], e\in[E-1]$ from \eqref{eq:disp_prob}.
\end{algorithmic}
}
\end{algorithm}

The preceding analysis establishes a system of self-consistency equations for the stationary quantities. Specifically, the dispatch probabilities $p_{ije}$ depend on $\rho_i$ and $p_i^e$ through \eqref{eq:disp_prob}, while \eqref{eq:Markov_transition}--\eqref{eq:pie} determine $\rho_i$ and $p_i^e$ from $\{p_{ije}\}$. This mutual dependence gives rise to a fixed-point system, which we solve via the Energy-Hypercube Iteration in Algorithm~\ref{algo:iterative_fixed_point}.

The feasibility of Algorithm~\ref{algo:iterative_fixed_point} is proved through 2 steps. Lemma~\ref{lemma:irreducibility} ensures that the Markov chain defined by \eqref{eq:Markov_transition} has a unique stationary distribution. Theorem~\ref{theorem:fixed_point} proves that Algorithm~\ref{algo:iterative_fixed_point} always has a fixed point using the mapping theorem.

\begin{lemma}
Assume that, for each unit $i$,
$
O_{ij1}>0,\; \forall j\in[J].
$
Then Equation~\eqref{eq:Markov_transition} defines an irreducible Markov chain on the finite state space
$
\mathcal S_i
=
\{(E,b_i)\}
\cup
\{(e,j): e=1,\dots,E-1,\ j=1,\dots,J\}.
$
Consequently, it admits a unique stationary distribution $\pi_i$, and
$
\pi_{(e,j)} > 0,\; \forall (e,j)\in \mathcal S_i.
$
\label{lemma:irreducibility}
\end{lemma}

\begin{theorem}[Existence of a fixed point]\label{theorem:fixed_point}
Assume that:  
(i) each demand node has a complete preference list over all units;  
(ii) $\lambda_{j1}>0$ for all demand nodes $j$;  
(iii) all travel, service, and recharging times are finite and strictly positive; and  
(iv) there exist constants $0<\underline{\rho}<\overline{\rho}<1$ and $0<\underline{p}<1/(E-\theta)$ such that the one-step update mapping induced by Algorithm~\ref{algo:iterative_fixed_point} maps
\[
\mathcal{K}
=
[\underline{\rho},\overline{\rho}]^N
\times
\prod_{i=1}^N \Delta_i(\underline{p})
\]
into itself, where
\[
\Delta_i(\underline{p})
=
\left\{
(p_i^{\theta+1},\dots,p_i^E)\in[\underline{p},1]^{E-\theta}:
\sum_{e=\theta+1}^E p_i^e=1
\right\}.
\]
Then Algorithm~\ref{algo:iterative_fixed_point} admits at least one fixed point in $\mathcal{K}$.
\end{theorem}

Exact stationary performances are generally unavailable in large-scale problems for hypercube models. The strongest analytical result is often the existence of a fixed point \citep{HuaSwersey2022FireMedics}. Nevertheless, Theorem~\ref{theorem:fixed_point} shows that, despite the additional complexity induced by the energy dimension, the proposed energy-aware hypercube approximation still admits a fixed point.

\subsection{Location--Zoning Problem}

Having developed a steady-state performance approximation for a given energy-constrained hypercube system, we now embed that performance model into a strategic design problem. Specifically, we determine where to open service bases and how to partition the service region into zones so as to optimize long-run system performance. The steady-state approximation is used to evaluate any given location--zoning design and therefore provides the objective value for the optimization model below.

Service-zone design is operationally important in our setting for two reasons. First, each ESV has a finite effective operating range due to battery limitations, so the service region must be partitioned in a way that allows units to reliably cover their assigned demand nodes. Second, long-distance dispatches increase travel time and reduce effective service capability. Restricting dispatches to designated zones helps concentrate service geographically and improve the effective utilization of limited battery resources.

Station location and zone design are coupled through a two-way dependence. Zoning decisions affect the value of opening a station by determining the demand and dispatch distances associated with its service zone, while station locations affect the quality of a zoning plan by determining how efficiently each zone can be served. This coupling motivates a joint optimization of station locations and service zones rather than a sequential design approach.

We consider a region with $M$ candidate station locations, from which $N$ sites are selected to open stations ($N<M$), where we assume that exactly one service unit is assigned to each opened station. The demand set consists of $J$ nodes, and the region is partitioned into $A$ disjoint service zones. Dispatch is allowed only within zones, thereby decoupling the operations of different zones.

Recall that each unit $i$ is assigned to a dedicated base station $b_i$. To simplify the network topology, we assume that each candidate station location is co-located with one of the demand nodes; that is, for each $i\in[M]:=\{1,\ldots,M\}$, there exists $j^\ast(i)\in[J]$ such that $b_i=j^\ast(i)$. Accordingly, in the design model we let $i\in[M]$ index candidate station locations, $a\in[A]$ index zones, and $j\in[J]$ index demand nodes. Let $\mathcal E:=\{1,\ldots,E-1\}$ denote the set of admissible energy-demand classes. Let binary variable $x_{ia}=1$ if a station is opened at candidate location $i$ and assigned to zone $a$, and let $x_{ia}=0$ otherwise. Let binary variable $d_{ja}=1$ if demand node $j$ is assigned to zone $a$, and let $d_{ja}=0$ otherwise. For each zone $a$, define the station vector $\bm{x}^a=(x_{ia})_{i\in[M]}$ and the demand-assignment vector $\bm{d}^a=(d_{ja})_{j\in[J]}$. Given $(\bm{x}^a,\bm{d}^a)$, the Energy-Hypercube Iteration in Section~\ref{sec:iter_alg} determines the steady-state dispatch probabilities $p_{ije}^a$. Hence, $p_{ije}^a$ is an implicit function of $(\bm{x}^a,\bm{d}^a)$ rather than an explicit decision variable in the design model.

We adopt the long-run profit rate as the objective. Let $r_{ije}$ denote the profit obtained when a type-$(j,e)$ demand is served by the unit located at station $i$. Then the objective is
$
\sum_{a,i,j,e} r_{ije}\lambda_{je}p_{ije}^a,
$
which is separable across zones.
More generally, because the objective is linear in $\{p_{ije}^a\}$, the same framework extends directly to any objective of the form $\sum_{a,i,j,e} w_{aije}p_{ije}^a$ with nonnegative weights $w_{aije}$, which encodes performance measures such as response time, service importance, or equity.

The resulting location--zoning problem (\textbf{LZP}) is formulated as 

{\SingleSpacedXI
\begin{subequations}\label{model:location-zoning}
\begin{align}
\textbf{LZP:}\quad \max\ & \sum_{a\in[A]}\sum_{i\in[M]}\sum_{j\in[J]}\sum_{e\in\mathcal E} r_{ije}\lambda_{je}p_{ije}^a \notag\\
\text{s.t.}\ 
& \sum_{a\in[A]} \sum_{i\in[M]} x_{ia} = N, && \label{LZconstr:station-num}\\
& \sum_{i\in[M]} x_{ia} \ge 1, && \forall a\in[A], \label{LZconstr:zone-has-station}\\
& x_{ia} \le d_{b_i a}, && \forall i\in[M],\, a\in[A], \label{LZconstr:station-zone-compat}\\
& \sum_{a\in[A]} d_{ja} = 1, && \forall j\in[J], \label{LZconstr:unique-zone-assign}\\
& \sum_{j':(j,j')\in \mathcal{A}} v_{jj'a}
  - \sum_{j'':(j'',j)\in \mathcal{A}} v_{j''ja}
  \ge d_{ja} - (J-A+1)\, h_{ja},
  && \forall j\in[J],\, a\in[A], \label{LZconstr:cont1}\\
& v_{jj'a} + v_{j'ja} \le (J-A)\, d_{ja}, && \forall j\in[J],\, j'\in[J],\, a\in[A], \label{LZconstr:cont2}\\
& \sum_{j\in[J]} h_{ja} = 1, && \forall a\in[A], \label{LZconstr:cont3}\\
& h_{ja} \le d_{ja}, && \forall j\in[J],\, a\in[A], \label{LZconstr:cont4}\\
& x_{ia} \in \{0,1\}, && \forall i\in[M],\, a\in[A], \label{LZconstr:x-binary}\\
& h_{ja},\, d_{ja} \in \{0,1\}, && \forall j\in[J],\, a\in[A], \label{LZconstr:hd-binary}\\
& v_{jj'a} \ge 0, && \forall j\in[J],\, j'\in[J],\, a\in[A]. \label{LZconstr:v-nonneg}
\end{align}
\end{subequations}
}

Constraint \eqref{LZconstr:station-num} fixes the total number of opened stations at $N$. Constraint \eqref{LZconstr:zone-has-station} requires each zone to contain at least one opened station, thereby ensuring local service capacity. Constraint \eqref{LZconstr:station-zone-compat} enforces consistency between station assignment and zone membership: if location $i$ is opened in zone $a$, then the demand node co-located with that station must also belong to zone $a$. Constraint \eqref{LZconstr:unique-zone-assign} assigns each demand node to exactly one zone. Constraints \eqref{LZconstr:cont1}--\eqref{LZconstr:cont4} impose zone contiguity through a flow-based representation \citep[see][]{shirabe2009districting}, where $v_{jj'a}$ denotes the auxiliary flow sent from node $j$ to node $j'$ within zone $a$, and $h_{ja}$ indicates whether node $j$ is selected as the sink node of zone $a$. Finally, constraints \eqref{LZconstr:x-binary}--\eqref{LZconstr:v-nonneg} specify integrality and nonnegativity requirements for the decision and auxiliary variables.

The \textbf{LZP} is computationally challenging for several reasons. First, it is a combinatorial design problem with binary location and zoning decisions, together with contiguity requirements on service zones. Second, because the system is deployed at the city scale, the number of candidate station locations, demand nodes, and feasible zone configurations can all be very large. Third, evaluating a feasible design requires solving the embedded system approximation, so the problem can be represented in an extended variable space as a mixed-integer nonlinear program (MINLP) with additional steady-state and fixed-point relations. At the same time, because dispatch is restricted within zones, the total profit is naturally additive over zone configurations, each of which combines a contiguous subset of demand nodes with the stations opened inside it. This additivity motivates the set-partitioning reformulation developed next, and we builds on it to develop a general algorithm framework that addresses both the combinatorial explosion of feasible zone configurations and the embedded nonlinear performance evaluation.

\section{Branch-Price-and-Evaluation Framework for Location--Zoning Problem}

This section develops a Branch-Price-and-Evaluation (BPE) framework for the \textbf{LZP}. A standard branch-and-price scheme is not directly applicable because the objective coefficient of a zone-selection column is implicit through external evaluation. That is, for a given zone-station configuration, the associated profit rate must be obtained by running the Energy-Hypercube Iteration in Section~\ref{sec:iter_alg}. To address this difficulty, we combine surrogate pricing with exact column evaluation.

The key idea is to solve the restricted master problem (RMP) at each branch-and-bound node through a nested \emph{column-generation-and-evaluation} (CGE) procedure. The inner loop performs \emph{surrogate pricing}: it uses tractable surrogate reduced cost to identify promising columns and add them to the RMP. The outer loop performs \emph{column evaluation}: once the inner loop terminates, it recovers the exact coefficients for the columns that currently determine the RMP solution and updates the RMP accordingly. The node-level procedure terminates when all such columns have been evaluated exactly. Branching is then applied to enforce integrality. In this way, BPE addresses both the combinatorial explosion in the number of feasible zone configurations and the implicit nonlinear structure of the objective. 

We reformulate the location--zoning problem as a set-partitioning master problem. Let $\mathbb{S}$ denote the set of all feasible zone configurations, where each configuration is represented by a pair $(S,I_S)$ with $S\subseteq [J]$ denoting a feasible zone and $I_S\subseteq S$ the set of stations opened in that zone. For each $(S,I_S)\in\mathbb{S}$, let $r_{S,I_S}$ denote the exact steady-state profit rate of configuration $(S,I_S)$, obtained by evaluating the induced energy-constrained hypercube system, and let $z_{S,I_S}$ be a binary variable that equals 1 if configuration $(S,I_S)$ is selected and 0 otherwise.
The resulting master problem is 

{\SingleSpacedXI
\begin{subequations}\label{MP}
\begin{align}
\textbf{MP:}\quad
\max\ & \sum_{(S,I_S)\in\mathbb{S}} r_{S,I_S}\, z_{S,I_S} \notag\\
\text{s.t.}\quad
& \sum_{\substack{(S,I_S)\in\mathbb{S}\\ j\in S}} z_{S,I_S} = 1,
&& \forall j\in[J],  \label{MP:cover}\\
& \sum_{(S,I_S)\in\mathbb{S}} |I_S|\, z_{S,I_S} = N,
&&  \label{MP:stations}\\
& \sum_{(S,I_S)\in\mathbb{S}} z_{S,I_S} = A,
&&  \label{MP:zones}\\
& z_{S,I_S}\in\{0,1\},
&& \forall (S,I_S)\in\mathbb{S}. \label{MP:binary}
\end{align}
\end{subequations}
}
The objective maximizes the total exact profit over the selected zone configurations, which is additive across zones. Constraint~\eqref{MP:cover} assigns each demand node to exactly one selected zone configuration. Constraint~\eqref{MP:stations} enforces the global station budget, and Constraint~\eqref{MP:zones} fixes the number of selected zones. Constraint~\eqref{MP:binary} imposes integrality on the column-selection variables.

Because $|\mathbb{S}|$ grows exponentially in $J$, we solve a restricted master problem (RMP) over a working set $\tilde{\mathbb{S}}\subseteq\mathbb{S}$. For each column $(S,I_S)\in\tilde{\mathbb{S}}$, let $\tilde r_{S,I_S}$ denote the coefficient currently used in the RMP. At any iteration,
\[
\tilde r_{S,I_S}=
\begin{cases}
r_{S,I_S}, & \text{if $(S,I_S)$ has already been evaluated exactly,}\\
\hat r_{S,I_S}, & \text{otherwise,}
\end{cases}
\]
where $\hat r_{S,I_S}$ is a surrogate coefficient used in the pricing step and constructed to satisfy $\hat r_{S,I_S}\ge r_{S,I_S}$. The specific surrogate constructions used in this paper are introduced in Sections~\ref{sec:pm-surrogate} and~\ref{sec:ua-surrogate}. The restricted master problem is

{\SingleSpacedXI
\begin{subequations}\label{RMP}
\begin{align}
\textbf{RMP:}\quad
\max\ & \sum_{(S,I_S)\in\tilde{\mathbb{S}}} \tilde r_{S,I_S}\, z_{S,I_S} \notag\\
\text{s.t.}\quad
& \sum_{\substack{(S,I_S)\in\tilde{\mathbb{S}}\\ j\in S}} z_{S,I_S} = 1,
   && \forall j\in[J], && (\pi_j) \label{RMP:cover}\\
& \sum_{(S,I_S)\in\tilde{\mathbb{S}}} |I_S|\, z_{S,I_S} = N,
   &&                 && (\pi_{J+1}) \label{RMP:stations}\\
& \sum_{(S,I_S)\in\tilde{\mathbb{S}}} z_{S,I_S} = A,
   &&                 && (\pi_{J+2}) \label{RMP:zones}\\
& z_{S,I_S}\ge 0,
   && \forall (S,I_S)\in\tilde{\mathbb{S}}. && \label{RMP:nonneg}
\end{align}
\end{subequations}
}
The dual variables associated with constraints~\eqref{RMP:cover}--\eqref{RMP:zones} are denoted by $\bm{\pi}=(\pi_1,\ldots,\pi_J,\pi_{J+1},\pi_{J+2})$, as indicated in parentheses on the right-hand side of the constraints.

Let $\mathbb{S}^\mathrm{u}\subseteq \tilde{\mathbb{S}}$ denote the set of columns whose coefficients have already been set to their exact values. For an optimal solution $\bar{\bm z}$ of the current \textbf{RMP}, define the \emph{active-column} set
$
\mathcal{C}(\bar{\bm z}) := \{(S,I_S)\in\tilde{\mathbb{S}}: \bar z_{S,I_S}>0\}.
$
The node-level CGE procedure alternates between (i) surrogate pricing over the current \textbf{RMP} and (ii) exact evaluation of columns in $\mathcal{C}(\bar{\bm z})\setminus\mathbb{S}^\mathrm{u}$. The node terminates when $\mathcal{C}(\bar{\bm z})\subseteq\mathbb{S}^\mathrm{u}$.

\subsection{Node-Level Column-Generation-and-Evaluation}

At each branch-and-bound node, exact pricing would require repeatedly evaluating $r_{S,I_S}$ via Energy-Hypercube Iteration, which is computationally prohibitive; hence, we solve the node via a nested CGE procedure. The inner loop solves the \textbf{RMP}, extracts the optimal dual vector $\bar{\bm\pi}$, and then solves a surrogate pricing problem to identify a column with positive surrogate reduced cost. This column is added to $\tilde{\mathbb{S}}$, and the inner loop continues until no such column exists. The outer loop then evaluates the exact coefficients of all active columns that have not yet been evaluated, replaces the corresponding surrogate coefficients $\hat r_{S,I_S}$ in the \textbf{RMP} with their exact coefficients $r_{S,I_S}$, and restarts the inner loop.

Under dual prices $\bar{\bm\pi}$, define the reduced cost of a candidate column $(S,I_S)$ as
\[
\tilde r_{S,I_S}-\sum_{j\in S}\bar\pi_j-|I_S|\bar\pi_{J+1}-\bar\pi_{J+2},
\]
which is exact when $\tilde r_{S,I_S}=r_{S,I_S}$, or a surrogate when $\tilde r_{S,I_S}=\hat r_{S,I_S}$. The validity of CGE requires the surrogate to upper-bound the exact coefficient of every feasible column.
Proposition~\ref{prop:p_median_bound} first provides such a surrogate through a $p$-median upper envelope. We later introduce a sharper uniform-acceptance surrogate, which is used in most of our computations.

\subsubsection{P-Median Surrogate and Pricing.}\label{sec:pm-surrogate}

We first construct a valid surrogate coefficient for each candidate column and then derive the associated pricing problem. 
The surrogate is motivated by the assignment structure of the classical $p$-median problem. For a fixed zone $S$ and a station budget $n$, if we assume that every unit is always available for dispatch, then the within-zone siting problem is to jointly choose $n$ open stations and assign each demand type $(j,e)$ to one open station so as to maximize the total reward, which is the same as a reward-maximizing $p$-median model. Our surrogate further specializes this idea to a given feasible column $(S,I_S)$: once the open-station set $I_S$ is fixed, we retain only the assignment part and assign each $(j,e)$ to the station in $I_S$ with the largest reward $r_{ije}$.

Accordingly, for a feasible column $(S,I_S)$, consider the induced assignment problem

{\SingleSpacedXI
\begin{align*}
\max\ &\sum_{i\in I_S}\sum_{j\in S}\sum_{e\in\mathcal E} \lambda_{je} r_{ije} y_{ije}\\
\textup{s.t.}\ &\sum_{i\in I_S} y_{ije}=1, &&\forall j\in S,\ e\in\mathcal E,\\
&y_{ije}\in\{0,1\}, &&\forall i\in I_S,\ j\in S,\ e\in\mathcal E.
\end{align*}
}
Its optimal value is
\[
\hat r^{\mathrm{PM}}_{S,I_S}
:=
\sum_{j\in S}\sum_{e\in\mathcal E}\lambda_{je}\max_{i\in I_S} r_{ije},
\]
which we call the \emph{$p$-median surrogate}, because it is exactly the assignment value induced by the $p$-median structure after fixing the open-station set.

\begin{proposition}
Assume that $r_{ije}\ge 0$ for all $i,j,e$, and that the exact dispatch probabilities satisfy
$p_{ije}\ge 0$
and
$\sum_{i\in I_S} p_{ije}\le 1$
for every $j\in S$ and $e\in\mathcal E$.
Then every feasible column $(S,I_S)$ satisfies
\[
r_{S,I_S}
\le
\hat r^{\mathrm{PM}}_{S,I_S}
=
\sum_{j\in S}\sum_{e\in\mathcal E}\lambda_{je}\max_{i\in I_S} r_{ije}.
\]
\label{prop:p_median_bound}
\end{proposition}
Proposition~\ref{prop:p_median_bound} shows that $\hat r^{\mathrm{PM}}_{S,I_S}$ is a valid column-specific upper bound on the exact coefficient. We therefore use it as the surrogate coefficient in pricing.

Given an optimal dual vector $\bar{\bm\pi}$ of the current \textbf{RMP}, the pricing problem can be written as a MILP using the binary zone-membership vector $\bm d=(d_j)_{j=1}^J$, station-location vector $\bm x=(x_j)_{j=1}^J$, and assignment variables $y_{ije}$, where $d_j=1$ if node $j$ belongs to the candidate zone, $x_j=1$ if a station is opened at node $j$, and $y_{ije}=1$ if demand class $(j,e)$ is assigned to station $i$. The surrogate-reduced-cost objective is
\begin{equation*}
\max\quad
\sum_{i\in[J]}\sum_{j\in[J]}\sum_{e\in\mathcal E}
\lambda_{je} r_{ije} y_{ije}
-\sum_{j\in[J]} \bar\pi_j d_j
-\bar\pi_{J+1}\sum_{i\in[J]} x_i
-\bar\pi_{J+2}.
\end{equation*}
The feasible region enforces assignment only to open stations, station--zone compatibility, the admissible station count for a single column, and the same flow-based contiguity structure used in \textbf{LZP}. The complete MILP formulation of \textbf{SPP} is given in Online Appendix~\ref{sec:app_spp}.

\subsubsection{Uniform-Acceptance Surrogate and Approximate Pricing.}\label{sec:ua-surrogate}

The $p$-median surrogate is easy to use in pricing but can be loose when service capacity is tight, because the $p$-median problem treats all demand as fully serviceable. To obtain a tighter and more informative surrogate, we introduce a uniform-acceptance (UA) surrogate that scales the $p$-median surrogate by an aggregate acceptance factor:
$
\hat r^{\mathrm{UA}}_{S,I_S}
:=
\eta_{S,I_S}\,\hat r^{\mathrm{PM}}_{S,I_S},
$
where $\eta_{S,I_S}\in[0,1]$ is a zone-level acceptance factor constructed from a conservative pooled-load approximation together with a piecewise-linear approximation of the Erlang-B blocking function. Relative to $\hat r^{\mathrm{PM}}_{S,I_S}$, this surrogate better captures the effect of blocking. More importantly, the resulting approximate surrogate pricing problem (\textbf{aSPP}) can also be reformulated as a MILP.

Unlike the $p$-median surrogate, however, the UA surrogate does not currently come with a theoretical guarantee that it upper-bounds the exact coefficient for every feasible column. We therefore use it as an empirically motivated surrogate. Nevertheless, across all instances in our computational study, the UA surrogate remains upper-bounding and is substantially tighter than the $p$-median surrogate, with a much stronger linear relationship to the exact coefficient. This stronger alignment makes the approximate pricing problem more likely to identify columns that are truly attractive under the exact coefficient, thereby improving the overall quality of the column generation process. The complete derivation, the full MILP formulation, and detailed empirical comparisons are deferred to Online Appendix~\ref{sec:app_ua_surrogate}.

\subsubsection{Exact Coefficient Evaluation and Updated-Column Exclusion.}

The surrogate pricing models above are used only to identify promising columns. Whenever the surrogate-pricing loop stops at $\bar{\bm z}$, CGE evaluates the exact coefficients of the columns in 
$\mathcal C(\bar{\bm z})\setminus \mathbb S^{\mathrm u}$ by the Energy-Hypercube Iteration in Section~\ref{sec:iter_alg}, and the RMP coefficient is replaced by
$
\tilde r_{S,I_S}\leftarrow r_{S,I_S}.
$
After this update, CGE checks whether 
all active columns have been updated; if not, it restarts surrogate pricing loop. 

The restart of surrogate pricing after exact coefficient evaluation requires an explicit mechanism to prevent updated columns from being regenerated. An updated column may still have positive surrogate reduced cost, because the surrogate coefficient remains an upper bound on the exact coefficient. To avoid this, after each exact-evaluation step we exclude every updated column $(S,I_S)\in\mathbb{S}^\mathrm{u}$ from subsequent pricing by imposing the Hamming-distance constraint
\[
1\le \|\bm d-\bm d^S\|_1+\|\bm x-\bm x^{I_S}\|_1,
\qquad \forall (S,I_S)\in\mathbb{S}^\mathrm{u},
\]
which can be written linearly as
\begin{align}
1
&\le \sum_{j=1}^J \bigl(-2d_jd^S_j+d_j+d^S_j\bigr)
   +\sum_{j=1}^J \bigl(-2x_jx^{I_S}_j+x_j+x^{I_S}_j\bigr),
&&\forall (S,I_S)\in\mathbb{S}^\mathrm{u}.
\label{constr:updated_r}
\end{align}

The full looping logic of CGE is given in Algorithm~\ref{algo:CGE} in Online Appendix~\ref{sec:app_bpe_logic}, and finite convergence follows from the node-level argument in Theorem~\ref{theorem:BPE_convergency}.

\subsection{Branch-Price-and-Evaluation Algorithm}\label{sec:BPE}

Embedding the node-level CGE procedure in branch-and-bound tree nodes yields our exact branch-price-and-evaluation (BPE) algorithm.
Algorithm~\ref{algo:CGE} in Online Appendix~\ref{sec:app_bpe_logic} gives the node-level routine. At each branch-and-bound node, CGE starts from the current \textbf{RMP} together with the branching decisions, and first removes all columns that are incompatible with those decisions. It then alternates between surrogate pricing by \textbf{SPP} or \textbf{aSPP} and exact coefficient evaluation. CGE terminates only when all active columns have been updated, at which point the node LP relaxation has been solved exactly.

The overall BPE framework, summarized in Algorithm~\ref{algo:BPE} in Online Appendix~\ref{sec:app_bpe_logic}, repeatedly applies CGE to active branch-and-bound nodes. After solving a node, the algorithm fathoms it if it is infeasible or dominated by the incumbent bound, updates the incumbent if the node solution is integral, and otherwise branches to create child nodes. We use Ryan--Foster pairwise branching whenever possible and fall back on branching on fractional $z_{S,I_S}$ when no effective pairwise branch remains. All branching decisions are enforced in pricing so that only branch-compatible columns can be generated; detailed branching constraints are deferred to Online Appendix~\ref{sec:app_branching}. Thus, CGE guarantees exact solution of every visited node relaxation, while branch-and-bound enforces integrality and optimality. The next theorem establishes finite convergence of the overall framework.

The next theorem establishes the global convergence of Algorithm~\ref{algo:BPE}. 
Its proof combines a node-level finite-convergence argument for Algorithm~\ref{algo:CGE},
obtained by specializing Theorem~2 of \citet{ZhangEtAl2023ScenarioReduction} to the LP
relaxation at each branch-and-bound node, with the standard correctness of branch-and-bound.

\begin{theorem}[Finite Convergence and Optimality of the BPE Framework]
\label{theorem:BPE_convergency}
Assume that, for every branch-and-bound node $n$, 
the surrogate pricing model upper-bounds the exact value of every node-feasible column. 
Then Algorithm~\ref{algo:BPE} terminates finitely and returns an optimal integer solution of \textbf{MP}.
\end{theorem}

The assumption is verified for \textbf{SPP} by Proposition~\ref{prop:p_median_bound}. The same conclusion applies to \textbf{aSPP} whenever its upper-bounding property holds for the instance under consideration.

The BPE framework extends branch-and-price to master problems with externally evaluated column coefficients. Its applicability relies on a surrogate that both bounds the true coefficients and remains solver-tractable; we provides such an envelope for the implicit energy-constrained hypercube coefficients. This suggests a general route for optimizing queueing systems with complex performance measures that need not be embedded in closed form in the pricing problem.

\subsection{A Scalable Heuristic Extension}\label{sec:BPE_heur}

The exact BPE algorithm provides an optimization framework with finite-convergence and optimality guarantees. For applications in larger city-scale instances, we further develop a scalable heuristic extension of BPE that is designed to generate high-quality solutions quickly. The heuristic preserves the column-generation-and-evaluation structure, while replacing exact tree search with a linear sequence of guided re-optimization at the root node. In each trial, we impose a small bundle of structural restrictions on the RMP. Varying these bundles across trials induces different search trajectories.

The heuristic also simplifies station-location decisions through aggregation. Instead of indexing columns by $(S,I_S)$, we index them by $(S,n)$, where $n$ is the number of stations assigned to zone $S$. The RMP therefore determines the zones and station counts, while the specific station set $I_S$ is selected by sampling or by a heuristic siting rule. Specifically, let
$
\mathcal I(S,n):=\{I_S\subseteq S:(S,I_S)\in\mathbb S,\ |I_S|=n\},
\;
r_{S,n}:=\max_{I_S\in\mathcal I(S,n)} r_{S,I_S}.
$
Under the aggregation
$
z_{S,n}:=\sum_{I_S\in\mathcal I(S,n)} z_{S,I_S},
$
we define \textbf{MP$'$} as an equivalent formulation of \textbf{MP}; a formal statement and proof are given in Online Appendix~\ref{sec:proof_prop_MP'_equiv}. Accordingly, the column-generation procedure is carried out on the corresponding \textbf{RMP$'$}. The heuristic approximation arises not from this aggregation itself, but from estimating $r_{S,n}$ by sampling or heuristic simplifications instead of exhaustively evaluating all $r_{S,I_S}$.

Overall, this heuristic extension retains the decomposition logic of BPE while avoiding branch-and-bound tree growth and exhaustive evaluation of station subsets. The numerical experiments below evaluate the computational performance of BPE and the proposed heuristic.

\section{Experimental Results}

We now turn from model development and algorithm design to numerical evaluation. We first validate the proposed exact and heuristic algorithms, and then use the resulting computational framework to examine the operational roles of energy modeling, zoning, and battery-capacity design.

\subsection{Algorithm Validation}

\begin{table*}[t]
\centering
\scriptsize
\caption{Performance comparison of enumeration, exact, and heuristic methods: optimal $R^\star$, computational time, and relative gaps (time in seconds).}
\label{tab:J9_results}

\setlength{\tabcolsep}{3.5pt}
\resizebox{\textwidth}{!}{%
\begin{tabular}{p{2.35cm}c
                rrr @{\hspace{4pt}}
                c@{\hspace{2pt}} c@{\hspace{2pt}}
                rrr @{\hspace{4pt}}
                r@{\hspace{2pt}}r}
\toprule
Problem & Instance &
\multicolumn{3}{c}{$R^\star$} &
\multicolumn{2}{c}{$\Delta R$ (\%)} &
\multicolumn{3}{c}{Time (s)} &
\multicolumn{2}{c}{$\Delta T$ (\%)} \\
\cmidrule(lr){3-5}\cmidrule(lr){6-7}\cmidrule(lr){8-10}\cmidrule(lr){11-12}
& &
Enum & Exact & Heur &
Exact--Enum & Heur--Enum &
Enum & Exact & Heur &
Exact--Enum & Heur--Exact \\
\midrule
\multirow{18}{2.55cm}{Synthetic}
& $N=2, A=2$ & 0.7251 & 0.7251 & 0.7163 & 0.00 & 1.22 & 58.88 & 150.12 & 211.78 & +154.94 & +41.08 \\
& $N=3, A=2$ & 1.0792 & 1.0792 & 1.0455 & 0.00 & 3.13 & 728.95 & 464.91 & 211.84 & -36.22 & -54.43 \\
& $N=4, A=2$ & 1.4162 & 1.4162 & 1.3649 & 0.00 & 3.62 & 2241.94 & 1194.53 & 216.71 & -46.72 & -81.86 \\
& $N=5, A=2$ & 1.7412 & 1.7412 & 1.6975 & 0.00 & 2.51 & 4006.55 & 2432.28 & 212.69 & -39.29 & -91.26 \\
& $N=6, A=2$ & 2.0608 & 2.0608 & 2.0176 & 0.00 & 2.10 & 5131.37 & 2710.29 & 186.13 & -47.18 & -93.13 \\
& $N=7, A=2$ & 2.3733 & 2.3733 & 2.3380 & 0.00 & 1.49 & 5532.63 & 2988.49 & 123.98 & -45.98 & -95.85 \\
& $N=8, A=2$ & 2.6855 & 2.6855 & 2.6167 & 0.00 & 2.56 & 5608.26 & 3255.77 & 211.96 & -41.95 & -93.49 \\
\addlinespace
& $N=3, A=3$ & 1.1900 & 1.1900 & 1.1900 & 0.00 & 0.00 & 35.64 & 138.40 & 211.26 & +288.30 & +52.65 \\
& $N=4, A=3$ & 1.5260 & 1.5260 & 1.5132 & 0.00 & 0.84 & 389.34 & 334.81 & 210.83 & -14.01 & -37.03 \\
& $N=5, A=3$ & 1.8636 & 1.8636 & 1.8436 & 0.00 & 1.07 & 1035.78 & 614.44 & 131.84 & -40.68 & -78.54 \\
& $N=6, A=3$ & 2.1965 & 2.1965 & 2.1839 & 0.00 & 0.57 & 1605.47 & 906.72 & 198.52 & -43.52 & -78.11 \\
& $N=7, A=3$ & 2.5196 & 2.5196 & 2.4951 & 0.00 & 0.98 & 1869.99 & 1055.05 & 211.30 & -43.58 & -79.97 \\
& $N=8, A=3$ & 2.8308 & 2.8308 & 2.7910 & 0.00 & 1.41 & 1940.72 & 1008.50 & 169.10 & -48.03 & -83.23 \\
\addlinespace
& $N=4, A=4$ & 1.5968 & 1.5968 & 1.5626 & 0.00 & 2.14 & 24.28 & 131.52 & 211.44 & +441.71 & +60.77 \\
& $N=5, A=4$ & 1.9330 & 1.9330 & 1.9046 & 0.00 & 1.47 & 247.09 & 288.24 & 211.34 & +16.65 & -26.68 \\
& $N=6, A=4$ & 2.2562 & 2.2562 & 2.2409 & 0.00 & 0.68 & 580.21 & 457.58 & 212.69 & -21.14 & -53.52 \\
& $N=7, A=4$ & 2.5640 & 2.5640 & 2.5272 & 0.00 & 1.44 & 801.82 & 611.14 & 213.81 & -23.78 & -65.01 \\
& $N=8, A=4$ & 2.8395 & 2.8395 & 2.8290 & 0.00 & 0.37 & 867.49 & 655.87 & 212.19 & -24.39 & -67.65 \\
\midrule
\multirow{2}{2.55cm}{Mobile charging}
& small & 0.4702 & 0.4702 & 0.4702 & 0.00 & 0.00 & 418.24 & 351.29 & 21.93 & -16.01 & -93.76 \\
& full  & -- & -- & 2.0779 & -- & -- & -- & -- & 2011.74 & -- & -- \\
\addlinespace
\multirow{2}{2.55cm}{Drone inspection}
& small & 0.4449 & 0.4449 & 0.4449 & 0.00 & 0.00 & 320.83 & 414.69 & 28.05 & +29.25 & -93.24 \\
& full  & -- & -- & 1.3468 & -- & -- & -- & -- & 1897.03 & -- & -- \\
\addlinespace
\multirow{2}{2.55cm}{Autonomous cleaning}
& small & 0.4705 & 0.4705 & 0.4705 & 0.00 & 0.00 & 2608.96 & 1618.61 & 57.16 & -37.96 & -96.47 \\
& full  & -- & -- & 1.5120 & -- & -- & -- & -- & 2147.93 & -- & -- \\
\bottomrule
\end{tabular}%
}

\vspace{1mm}
\parbox{0.97\textwidth}{\footnotesize
\textit{Notes.} Synthetic instances use $J=9$. For each real-case problem, the small and full layers use $J=9$ and $J=36$, respectively; only Heur is run on the full layer, so unavailable entries are marked ``--''. The real-case parameter sets are constructed from Toronto spatial data: mobile charging (small: $A=3$, $N=6$; full: $A=4$, $N=27$), inspection drone (small: $A=3$, $N=6$; full: $A=4$, $N=9$), and autonomous cleaning robot (small: $A=3$, $N=6$; full: $A=4$, $N=12$). For a pair of methods $(m_1,m_2)$, $\Delta R_{m_1-m_2}=100(R_{m_1}-R_{m_2})/R_{m_2}$ and $\Delta T_{m_1-m_2}=100(T_{m_1}-T_{m_2})/T_{m_2}$; negative $\Delta T$ indicates a speedup.}
\end{table*}

This subsection evaluates the solution quality and computational performance of three methods for the location--zoning problem: full enumeration (Enum), the exact BPE algorithm with aSPP pricing (Exact), and the BPE-based heuristic (Heur). For each method, we report the resulting objective value and the solution time. Under full enumeration, we evaluate the profit contribution $r_{S,I_S}$ for every feasible zone configuration $(S,I_S)$ using the stationary approximation and then solve the resulting master problem exactly; this yields the true optimal solution. We test the three methods on synthetic and real-data-based instances. The synthetic instances fix $J=9$ demand nodes and vary $(A,N)$, the numbers of service zones and service units, respectively. For the three real-case applications, we consider a small layer ($J=9$) and a full layer ($J=36$); for the full layer, only the heuristic results are reported.

The three real-case settings---mobile charging, drone inspection, and autonomous cleaning---fit naturally into the modeling framework developed in this paper. In all of them, service is delivered under a server-to-customer operating mode: a mobile unit is initially stationed at a base, travels to the demand location to provide service, and, once its remaining energy drops below a prescribed threshold, returns to a base for replenishment before re-entering the system. The main application-specific difference lies in how service consumes energy: in mobile charging, energy is transferred to an EV during service; in drone inspection, energy is consumed while performing the inspection task on site; and in autonomous cleaning, energy is consumed while cleaning for a task-dependent duration. These differences affect the operational interpretation of service but do not alter the common energy-constrained service structure. Hence, the same energy-constrained hypercube model and location--zoning structure can be applied across these applications.

For the three cases, we calibrate the case-specific parameters using public demographic data and representative product specifications. Baseline arrivals are scaled by FSA-level population counts from the 2021 Census \citep{statcan2021FSAPopulation}. In the mobile-charging case, an on-site discharge rate is set to 48 kW; this is an upper-end but commercially observed setting consistent with the typical 40--50 kW output range \citep{siemensHelioxMobile40,lincolnVelion50Mobile,kempowerMovableCharger}. Each mobile charger carries about 160 kWh \citep{freewireBoostCharger150}, divided into $E=10$ energy levels. The travel speed is set to 16 km/hour as a conservative baseline speed based on congested urban traffic in Toronto \citep{tomtomTorontoTraffic2025}. The drone-inspection and autonomous-cleaning cases use comparable arrival baselines, and their parameters are benchmarked against typical product data \citep{djiMavic3EnterpriseSpecs,boschungUrbanSweeperS20}. Detailed numerical values and source are reported in Online Appendix~\ref{sec:app_params}.

Table~\ref{tab:J9_results} summarizes the numerical results. On all instances for which full enumeration is computationally available, Exact reproduces the enumeration optimum. This also provides empirical support for the aSPP-based implementation, although the formal global-optimality guarantee in Section~\ref{sec:ua-surrogate} applies to SPP, or more generally to surrogate pricing schemes satisfying $\hat r>r$. Exact is usually faster than Enum on larger synthetic instances, while enumeration remains competitive only when the BPE overhead dominates on very small cases. Heur achieves a much stronger scalability--quality tradeoff: its synthetic optimality gap never exceeds $3.62\%$ and is often below $2\%$, and it matches the Exact/Enum solution on all three small real-case applications. For the full real-case layers, where Enum and Exact are computationally impractical, Heur remains tractable. The strong performance of Heur is consistent with its design in Section~\ref{sec:BPE_heur}. Overall, these results support Exact when an exact solution is required and Heur when scalability is the primary concern.

\subsection{Why Energy Modeling Matters}

Energy modeling affects both operational reliability and strategic design. Operationally, an energy-agnostic model may recommend dispatching a unit that is spatially available but energy-infeasible, leading to unreliable system performance. Strategically, omitting energy dynamics can misestimate the performance of station configurations and bias location decisions. We therefore quantify these decision-quality losses and identify the regimes in which energy constraints are either indispensable or safely approximated away.

\subsubsection{Decision Quality of Energy-Aware and Energy-Agnostic Models.}

To strategically isolate the impact of energy considerations, we compare four models that produce base-location decisions:
(i) the proposed energy-aware hypercube model (energy-constrained hypercube),
(ii) an energy-agnostic hypercube model with a profit-rate objective (labeled ``max Profit Rate''),
(iii) an energy-agnostic hypercube model with a response-time objective (labeled ``min RT''),
and (iv) a $p$-median variant with a profit objective that assigns each demand node to its nearest chosen base. 

We consider a no-zoning setting and optimize only the base-location vector $x\in\{0,1\}^J$ with $\sum_{j=1}^J x_j=N$.
For a fixed instance ($J=9$, $N=6$) and a demand scaling factor $\alpha$ (multiplying all arrival rates), we enumerate all $\binom{J}{N}$ feasible base-location decisions and evaluate each by discrete-event simulation. We take the resulting long-run profit rate $R^{\text{sim}}(x;\alpha)$ as ground truth. For each model $m$ and each $\alpha$, we compute the optimal decision
$x^{m}(\alpha)\in\arg\max_x \widehat{R}^{\,m}(x;\alpha)$ (where $\widehat{R}^{\,m}$ is the model's objective value), and assess it by its simulation performance $R^{\text{sim}}(x^{m}(\alpha);\alpha)$.

\begin{figure}[!ht]
    \centering
    \includegraphics[width=0.60\linewidth]{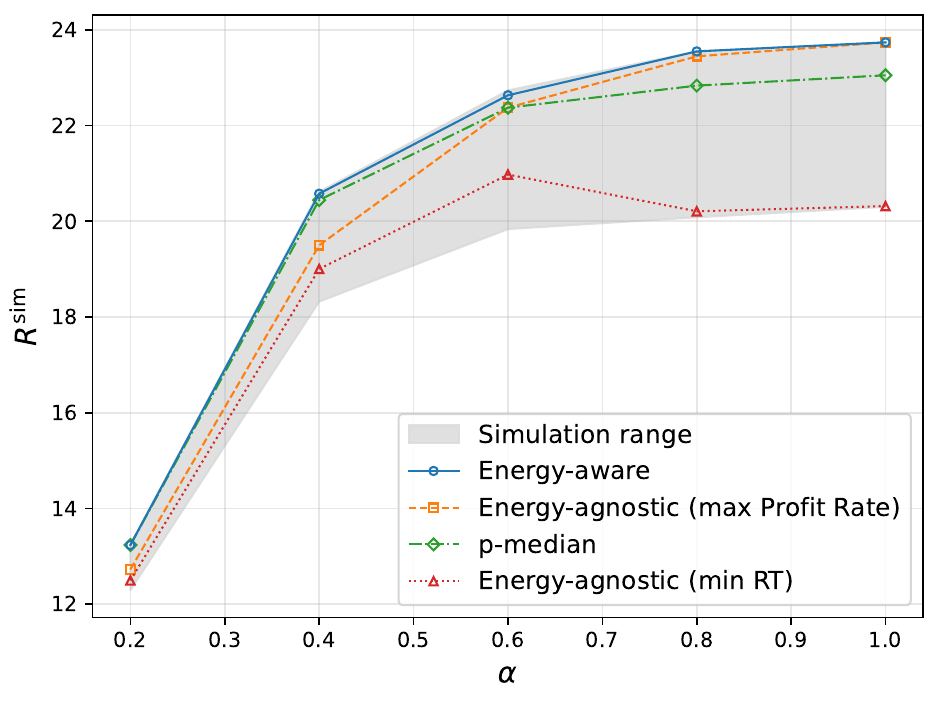}
    \caption{Simulation-evaluated decision quality of four location models for the no-zoning instance with $J=9$ and $N=6$, as the demand scaling factor $\alpha$ varies. For each model $m$, the colored curve plots $R^{\text{sim}}(x^{m}(\alpha);\alpha)$, i.e., the long-run profit rate in simulation of the location decision selected by that model. The shaded band shows the full simulation range $\left[\min_x R^{\text{sim}}(x;\alpha),\max_x R^{\text{sim}}(x;\alpha)\right]$ over all feasible base-location decisions, and its upper boundary corresponds to the simulation-optimal decision.}
    \label{fig:loc_decision_quality_4_models}
\end{figure}

Figure~\ref{fig:loc_decision_quality_4_models} delivers two main insights.
First, the energy-aware hypercube yields the best decisions across the full range of system loads:
its curve consistently lies closest to the upper boundary of the simulation range, indicating that the proposed energy-constrained hypercube captures the key operational tradeoffs that drive profitable siting decisions under energy constraints.
Second, the two energy-ignorant benchmarks exhibit a load-dependent ranking.
At low load (small $\alpha$), the simulation range is relatively narrow because most demand is fulfilled under nearly any feasible siting decision; as a result, differences in realized profit are driven mainly by travel cost rather than by differences in service revenue. The $p$-median decision is then competitive and can outperform the energy-agnostic hypercube with a profit-rate objective.
As $\alpha$ increases, the simulation range widens because demand fulfillment becomes increasingly constrained by unit unavailability. The energy-agnostic hypercube then overtakes the static-assignment $p$-median benchmark because it captures the fact that the realized fulfillment  falls below one at higher loads, whereas the latter does not; consequently, the two curves cross. 
The ``min RT'' benchmark is included as a classical energy-agnostic hypercube baseline from the literature.

Overall, different modeling assumptions lead to materially different siting decisions, and explicitly modeling energy states yields consistently better ones.
In the next subsection, we trace the mechanism behind the underperformance of the energy-agnostic models by decomposing the observed service shortfalls into busy-blocking versus energy-blocking mechanisms.

\subsubsection{The Role of Energy.}

Using the same no-zoning setting as in the previous subsection, we operationally isolate the role of energy constraints by considering three energy-demand regimes for incoming requests. We consider three energy-demand regimes for incoming requests: Mix-L (low-energy-demand mix), Mix-M (moderate), and Mix-H (high), which shift probability mass toward larger energy requirements $e$ (interpretable as larger charging amounts in a mobile-charging application). For each regime, we scale the total arrival rates by a load factor $\alpha\in\{0.2,0.4,0.6,0.8,1.0\}$.

To quantify the consequence of ignoring energy, we use the false fulfillment ratio (FFR). Let $u^{\text{model}}_{je}$ and $u^{\text{sim}}_{je}$ denote the model-predicted and simulation-realized fulfillment rates for type-$(j,e)$ demand.
We define
\[
\mathrm{FFR} \;=\;
\frac{\sum_{j,e}\lambda_{je}\max\{u^{\text{model}}_{je}-u^{\text{sim}}_{je},0\}}
{\sum_{j,e}\lambda_{je}u^{\text{model}}_{je}},
\]
which measures the fraction of the model's ``promised'' service workload that is not actually deliverable. We further decompose $\mathrm{FFR}$ by attributing the over-predicted workload to the root cause observed in simulation: busy-blocking component (no idle unit exists) versus energy-blocking component (an idle unit exists but is not energy-feasible).

\begin{figure}[t]
    \centering
    \includegraphics[scale=0.4]{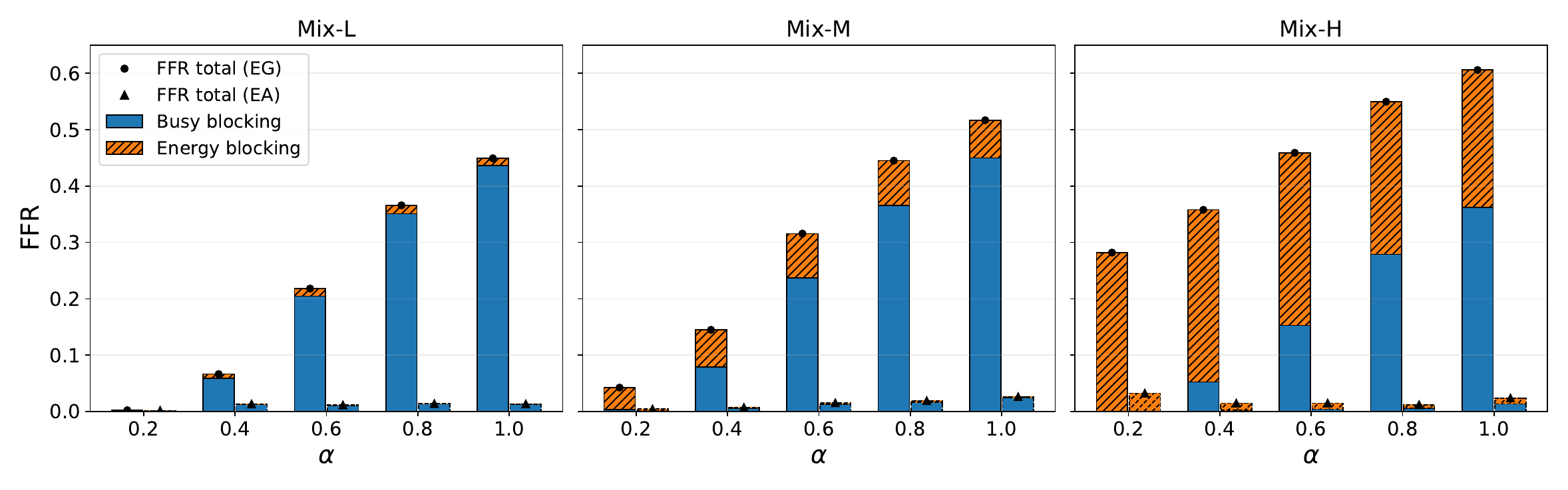}
    \caption{False fulfillment ratio (FFR) and its decomposition for the energy-agnostic (EG) and energy-aware (EA) hypercube models. Each panel corresponds to one energy-demand regime. The stacked bars decompose FFR into busy-blocking and energy-blocking components, and the marker reports total FFR.}
    \label{fig:ffr_decomp}
\end{figure}

Figure~\ref{fig:ffr_decomp} shows that explicit energy modeling sharply reduces false promises. Across all demand regimes, the energy-aware model (EA) delivers much smaller FFR than the energy-agnostic benchmark (EG). At $\alpha=1.0$, EG's FFR ranges from $0.449$ to $0.606$ across three regimes, whereas EA keeps it below $0.03$ throughout.  Additional distributional evidence on the estimation errors of $u_{je}$ is reported in Online Appendix~\ref{sec:app_role_energy}.

The decomposition also clarifies the mechanism. Under high-energy-demand mix and light load, EG's error is driven almost entirely by energy infeasibility rather than busy blocking: in Mix-H at $\alpha=0.2$, EG yields $\mathrm{FFR}=0.282$, of which $0.281$ is energy-driven. As load increases, busy blocking becomes more important in all regimes, but energy infeasibility remains a material source of service loss in the high-energy regime: in Mix-H at $\alpha=1.0$, EG still has an energy-blocking component of $0.244$.

These results show that ignoring energy does not merely reduce average accuracy; it misrepresents the source of service loss. When high-energy requests are prevalent, the presence of idle units is not by itself informative about actual service capability, because some of those units may be unable to serve the request. More broadly, reliable planning in energy-constrained service systems requires modeling both busy blocking and energy feasibility explicitly.

\subsection{Why Zoning Can Help, and When It Hurts}

\begin{figure}[t]
\centering
\includegraphics[width=0.60\linewidth]{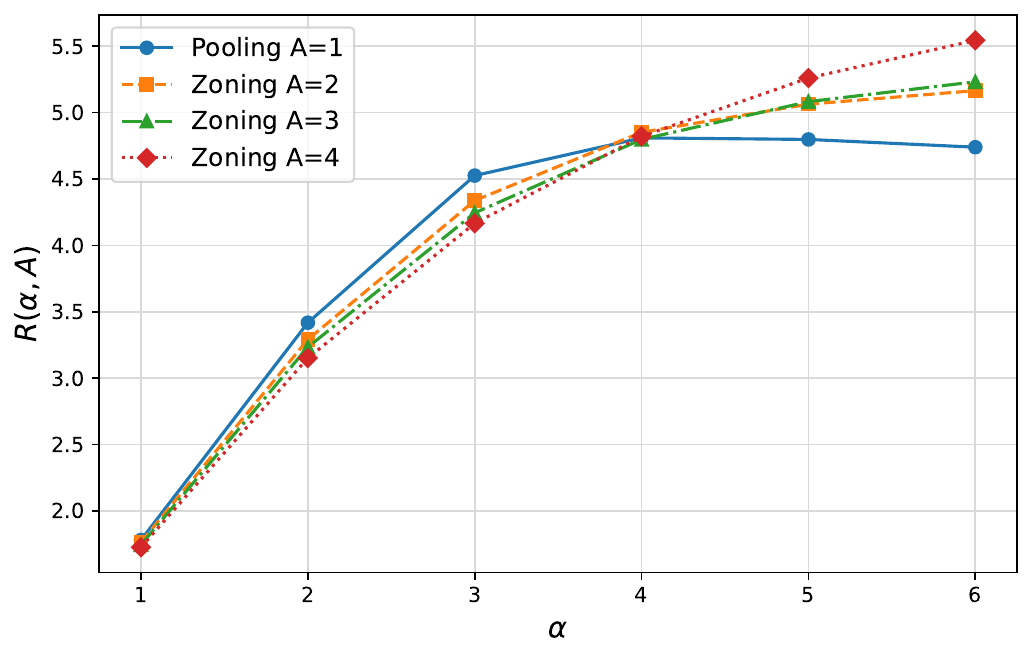}
\caption{Profit rate $R(\alpha,A)$ (in CAD/min) across load levels and zoning intensities. Zoning is detrimental under light load but beneficial under heavy load, with the transition occurring at intermediate load.}
\label{fig:profit_rate_vs_alpha}
\end{figure}

To understand when zoning improves performance in the mobile-charging system and when it becomes counterproductive, we conduct a set of computational experiments on a connected Toronto sub-instance with $J=27$ demand nodes and a fixed fleet size of $N=24$. We vary the load multiplier $\alpha\in\{1,2,3,4,5,6\}$ and the zoning intensity $A\in\{1,2,3,4\}$, where $A=1$ corresponds to pooling and $A>1$ imposes progressively finer geographic partitioning. For each $(\alpha,A)$, we report the system-wide profit rate $R(\alpha,A)$.

To diagnose the mechanisms behind the profit patterns, we use two simulation-based indicators under the same optimized deployment and assignment decisions. The first is travel time per completed service,
\[
\mathrm{TPS}(\alpha,A)
=
\frac{T_{\text{travel}}(\alpha,A)}{N_{\text{served}}(\alpha,A)},
\]
where $T_{\text{travel}}(\alpha,A)$ denotes the total travel time incurred by the fleet under configuration $(\alpha,A)$, including travel to customers and required return travel, and $N_{\text{served}}(\alpha,A)$ denotes the number of completed services. The second is the zoning-block share of unserved requests,
\[
\mathrm{ZB}(\alpha,A)
=
\frac{B(\alpha,A)}{U(\alpha,A)},
\]
where $U(\alpha,A)$ is the total number of unserved requests, and $B(\alpha,A)$ is the number of unserved requests for which there exists at least one globally idle and battery-feasible unit at the arrival epoch. A high value of $\mathrm{ZB}(\alpha,A)$ indicates that a substantial share of lost demand is caused by the zoning constraint, in the sense that these requests could potentially have been matched with available units in the absence of zoning, rather than by a system-wide shortage of such units. This measure should be interpreted as a diagnostic indicator rather than a literal recovery rate: a request counted in $B(\alpha,A)$ would not necessarily be served if zoning were removed. Nevertheless, the variation of $\mathrm{ZB}(\alpha,A)$ across $\alpha$ and $A$ is informative about the relative importance of demand loss caused by zoning.

Figure~\ref{fig:profit_rate_vs_alpha} shows a clear load-dependent reversal in the value of zoning. Under light load, pooling dominates: zoning lowers profit, and the loss becomes larger as the partition becomes finer. Under heavy load, the ranking reverses: zoning outperforms pooling, and stronger zoning becomes more attractive. The transition occurs in the intermediate-load region, where the benefit from shorter dispatches begins to offset the demand loss caused by zoning. Importantly, the high-load profit gains are not primarily driven by higher fulfillment. Instead, they arise because zoning reduces travel time per completed service and therefore lowers the travel cost per completed job, increasing the net profit earned per service.

\begin{figure}[t]
\centering
\includegraphics[width=0.60\linewidth]{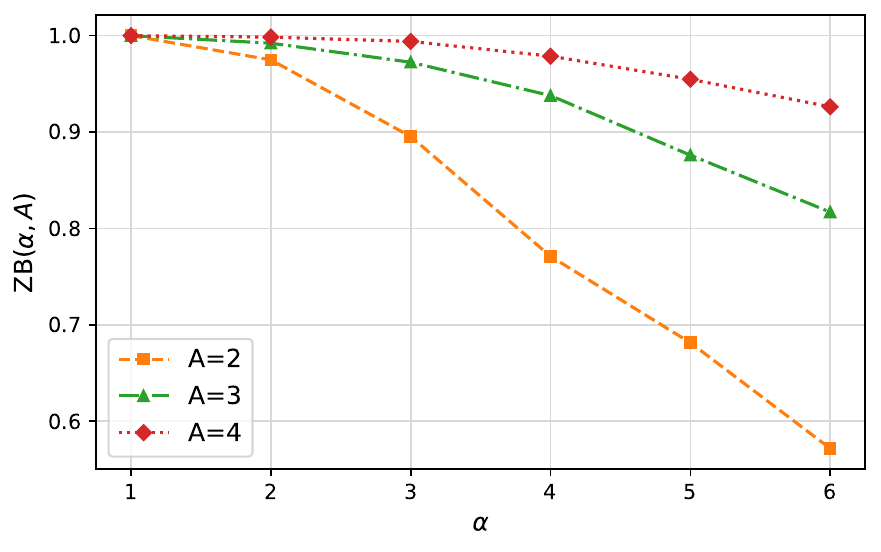}
\caption{Zoning-block share of unserved requests, $\mathrm{ZB}(\alpha,A)$, across load levels and zoning intensities. Fragmentation is most pronounced under light load and weakens as load increases, although it remains stronger under finer partitions.}
\label{fig:zoning_block_share}
\end{figure}

The mechanism is consistent with Figure~\ref{fig:zoning_block_share} and the supplementary travel-burden evidence reported in Online Appendix~\ref{sec:app_zoning_mechanism}. At low load, TPS is already modest under pooling, so zoning has limited room to generate meaningful travel savings. By contrast, ZB is extremely high , indicating that many lost requests are caused by zoning-induced capacity isolation. In this regime, zoning primarily reduces matching flexibility and therefore lowers profit.

As load increases, the mechanism changes. Under pooling, TPS rises sharply with load, indicating that a growing share of operational effort is spent on inefficient dispatches. Zoning counteracts this effect by localizing service territories and reduces the travel burden per service; this reduction becomes stronger as zoning intensity increases. At the same time, ZB declines with load, especially under moderate zoning, implying that zoning-induced demand loss becomes less important when the system is busy. Intuitively, once demand is high, the main operational challenge is no longer to preserve global flexibility at all times, but to avoid spending excessive effort moving units across space. In that regime, the travel-side savings from zoning outweigh the demand loss that zoning still creates, so the net effect on profit becomes positive.

Taken together, the evidence suggests that zoning creates a fundamental trade-off between matching flexibility and shorter dispatches. Pooling preserves the ability to use available units and is therefore more effective when demand is light. Zoning restricts this flexibility, but it shortens dispatch distances and lowers travel time per service when demand is dense. These findings indicate that zoning is an important design consideration in our problem. In particular, for city-scale service regions, a model that ignores zoning can lead to high travel-related operating costs. This motivates our formulation of a joint location--zoning design problem rather than a pure location problem.

\subsection{How Managers Should Jointly Design Battery Capacity and Zoning}\label{sec:joint_battery_zoning}

Battery capacity is a central but non-monotone design lever in ESV systems. Larger batteries can reduce energy-related interruptions and expand the feasible service range, but they also interact with unit cost, charging time, and service-region design \citep{liu2026mobileChargingRobots,cheng2020droneRoutingEnergy,puduCC1Pro}. The marginal value of capacity therefore depends on zoning: a well-designed zoning structure may substitute for excessive battery investment, whereas a poorly matched spatial design may leave additional capacity underutilized.
We next examine how managers should coordinate battery technology and spatial design using the full-layer mobile-charging instance reported in Table~\ref{tab:J9_results}. Online Appendix~\ref{sec:app_joint_design} reports the revenue and its increments for all $(E,\alpha,A)$ combinations. Three patterns emerge.

First, the main zoning pattern is unchanged across the tested battery range: zoning is less attractive under light demand but becomes increasingly valuable as demand rises. In other words, there is a robust load-dependent reversal in zoning, with pooling ($A=1$) preferred when $\alpha=1$ and the tightest zoning design ($A=4$) preferred for $\alpha=4,7,10$ at every tested battery level.

Second, larger battery capacity and tighter zoning are partial substitutes. As $E$ increases, the revenue gap between weaker and stronger zoning compresses systematically. For example, at $\alpha=4$, the revenue advantage of $A=4$ over $A=1$ declines from $0.4483$ when $E=10$ to $0.3038$ when $E=22$, indicating that both levers partly mitigate the same inefficiency associated with long and energy-stressing dispatches.

Third, battery upgrades matter mainly at moderate and high load, but spatial design is the stronger lever across all tested parameter combinations. For example, at $\alpha=7$, moving from $A=1$ to $A=2$ at $E=10$ raises profit rate by $0.6552$, whereas increasing $E$ from $10$ to $22$ under $A=1$ raises profit rate by only $0.2325$. More generally, for moderate and high load, a one-level increase in zoning dominates expanding battery capacity from $E=10$ to $E=22$ within the same zoning level. 

A notable implication is that more battery capacity is not uniformly beneficial. At $\alpha=1$, increasing battery capacity slightly reduces revenue under every zoning design, indicating that under sparse demand, larger batteries may delay replenishment and reduce fleet readiness. The reason is that when requests arrive infrequently, additional range is less likely to be converted into immediate service, so a longer deployment cycle can reduce effective availability rather than improve it. Under heavier load, by contrast, additional capacity is more readily used to serve consecutive requests before units need to return for recharge.

These findings yield three managerial implications. First, battery expansion is valuable primarily when demand is sufficiently high that energy constraints materially limit service capacity. Second, because battery capacity and tighter zoning alleviate similar operational inefficiencies, they should be viewed as substitute levers rather than independent design choices. Third, zoning is the more powerful lever. Therefore, managers should first determine the zoning structure that matches the demand regime, and then evaluate the marginal value of battery expansion under that spatial design.

\section{Conclusion}

This paper develops an integrated analytical and optimization framework for energy-constrained spatial service systems. Motivated by emerging ESV applications, such as mobile EV charging, drone-based inspection, and autonomous field service, we show that energy constraints are not merely a reduction in effective service capacity. They reshape server availability, dispatch feasibility, and ultimately the logic of spatial system design. Capturing these mechanisms is therefore essential for credible long-run performance evaluation and service-system design.

Methodologically, the paper connects descriptive and prescriptive analytics for this class of systems. On the descriptive side, we extend the classical hypercube model to incorporate battery-state dynamics and energy-driven replenishment, providing a tractable representation of steady-state performance in energy-constrained spatial service systems. On the prescriptive side, we formulate a joint location--zoning design problem and solve it through a Branch-Price-and-Evaluation framework, which provides a general solution approach for large-scale optimization with embedded queueing-based performance evaluation and nonlinear system interactions.

The numerical analysis shows that modeling energy explicitly is not a secondary refinement but a prerequisite for high-quality planning. Energy-agnostic models can generate operationally false service promises and strategically inferior siting decisions because they misrepresent the mechanisms through which energy infeasibility and server unavailability jointly reduce system performance. The results also show that zoning should not be viewed as uniformly beneficial or harmful: its value depends on the demand regime. Finally, the joint battery--zoning analysis suggests that battery improvement and tighter zoning partly address the same inefficiency, but that spatial design is often the more powerful managerial lever. This implication is further sharpened by the counterintuitive finding that larger batteries may delay replenishment and reduce fleet readiness under sparse demand. Therefore, the managerial priority should be to first match the zoning structure to the demand regime and then evaluate the marginal value of additional battery capacity. 

More broadly, this paper positions ESV operations as a class of mobile service systems in which energy and space must be designed jointly. Beyond the focal mobile-charging application, this framework may also be adapted to other emerging ESV applications, such as resident underwater robots and agricultural field robots. The proposed energy-constrained hypercube model and Branch-Price-and-Evaluation framework provide one step toward a planning theory that addresses not only where to place resources, but also how to organize mobile capacity whose availability evolves with both service demand and replenishment cycles. Future research can extend this framework by incorporating dynamic repositioning, time-varying demand, and interactions with power-grid constraints. These extensions would further support the design of reliable, scalable, and adaptive ESV networks in increasingly electrified and automated service environments.




\putbib[reference]
\end{bibunit}






\ECSwitch 

\ECHead{E-Companion for ``On-Demand Service Zone Design for Energy-Constrained Spatial Queueing System''}

\begin{bibunit}[pomsref]
\section{Notation}
\label{sec:notation}

\refstepcounter{table}
\label{tab:notation}
\addcontentsline{lot}{table}{\protect\numberline{\thetable}{Summary of notation.}}

\begingroup
\setlength{\LTpre}{0pt}
\setlength{\LTpost}{0pt}
\setlength{\tabcolsep}{4pt}
\renewcommand{\arraystretch}{1.12}

{\centering
\EGT\TableCaptionFontStyle
{\TableNameFontStyle Table\hskip0.3em\thetable\kern16pt}Summary of notation.\par
}

\vspace{5pt}

\SingleSpacedXI
\footnotesize

\begin{longtable}{@{}L{0.24\textwidth}L{0.70\textwidth}@{}}
\toprule
Symbol & Description \\
\midrule
\endfirsthead

\toprule
Symbol & Description \\
\midrule
\endhead

\midrule
\multicolumn{2}{r}{\textit{Continued on next page}}\\
\endfoot

\bottomrule
\endlastfoot

\multicolumn{2}{@{}l}{\textbf{Indices}}\\
\addlinespace[2pt]
$i$ & Index of an ESV unit or a candidate station location. \\
$j$ & Indices of demand nodes. \\
$j^\ast(i)$ & Demand node co-located with candidate station location $i$. \\
$b_i$ & Dedicated base station of unit $i$. \\
$\gamma_{jk}$ & Index of the $k$th preferred unit in the preference list of demand node $j$. \\
$e$ & Indices of discrete energy levels. \\
$a$ & Index of a service zone. \\

\addlinespace[4pt]
\multicolumn{2}{@{}l}{\textbf{Sets}}\\
\addlinespace[2pt]
$[K]$ & Set $\{1,\ldots,K\}$ for a positive integer $K$. \\
$\mathcal E$ & Set of admissible energy-demand classes, $\mathcal E=\{1,\ldots,E-1\}$. \\
$\mathcal S_i$ & State space of unit $i$, $\mathcal S_i=\{(E,b_i)\}\cup\{(e,j):e=1,\ldots,E-1,\ j=1,\ldots,J\}$. \\
$\mathcal A$ & Set of adjacency arcs used to impose spatial contiguity. \\
$S, I_S$ & Set of demand nodes forming a feasible service zone and set of stations opened in zone $S$, respectively. \\
$\mathcal I(S,n)$ & Set of feasible station subsets of size $n$ inside zone $S$. \\
$\mathbb S, \tilde{\mathbb S}, \mathbb S^\mathrm{u}$ & Full feasible column set, current restricted-master column set, and set of updated columns, respectively. \\
$\mathcal C(\bar{\bm z})$ & Active-column set under solution $\bar{\bm z}$, $\mathcal C(\bar{\bm z})=\{(S,I_S)\in\tilde{\mathbb S}:\bar z_{S,I_S}>0\}$. \\

\addlinespace[4pt]
\multicolumn{2}{@{}l}{\textbf{Basic system parameters}}\\
\addlinespace[2pt]
$J$ & Number of demand nodes. \\
$N$ & Number of ESV units or opened stations. \\
$M$ & Number of candidate station locations. \\
$E$ & Maximum effective service-energy capacity, measured in discrete energy levels. \\
$A$ & Number of service zones. \\
$\theta$ & Battery threshold below which a unit returns to its base for replenishment. \\
$\lambda_{je}$ & Arrival rate of type-$(j,e)$ demand. \\
$\lambda_j$ & Total arrival rate at node $j$, $\lambda_j=\sum_{e\in\mathcal E}\lambda_{je}$. \\
$\tau^\mathrm{t}_{jj'}$ & Travel time from node $j$ to node $j'$. \\
$\tau^\mathrm{s}_{e}$ & On-site service time for a demand requiring $e$ energy levels. \\
$\tau^\mathrm{rt}_{ij}$, $\tau^\mathrm{rt}_{j}$ & Return travel time from node $j$ to the base of unit $i$; the unit index is suppressed when fixed. \\
$\tau^\mathrm{rc}_{e}$ & Recharging or replenishment time from battery level $e$ to full charge $E$. \\
$r_{ije}$ & Profit obtained when a type-$(j,e)$ demand is served by the unit located at station $i$. \\

\addlinespace[4pt]
\multicolumn{2}{@{}l}{\textbf{Energy-constrained hypercube approximation quantities}}\\
\addlinespace[2pt]
$p_{ije}$ & Stationary probability that unit $i$ is dispatched to serve a type-$(j,e)$ demand. \\
$\rho_i$ & Utilization of unit $i$, i.e., the long-run probability that unit $i$ is busy serving or replenishing. \\
$p_i^e, p_i^{>e}, p_i^{\le e}$ & Conditional probabilities for idle unit $i$ having battery level equal to $e$, above $e$, and at most $e$, respectively, with $p_i^{>e}=p_i^{e+1}+\cdots+p_i^E$ and $p_i^{\le e}=1-p_i^{>e}$. \\
$\rho_i^e$ & Probability that unit $i$ is unavailable to a type-$(j,e)$ demand, $\rho_i^e=\rho_i+(1-\rho_i)p_i^{\le e}$. \\
$\bar\rho^e$ & Average unavailability probability for energy-demand class $e$, $\bar\rho^e=N^{-1}\sum_{i=1}^N\rho_i^e$. \\
$Q(N,\bar\rho,k)$ & Hypercube correction factor used to approximate dependence among unit states. \\
$O_{ije}$ & Offer rate of type-$(j,e)$ demands to unit $i$. \\
$O_{ie}$ & Total offer rate to unit $i$ for demands requiring $e$ energy levels, $O_{ie}=\sum_j O_{ije}$. \\
$\mathbf P^i$ & Embedded state-to-state transition matrix for unit $i$. \\
$\mathbf P_{(e,j)(e',j')}$ & Transition probability from state $(e,j)$ to state $(e',j')$ for a fixed unit. \\
$\bar\tau_e^\mathrm{w}$ & Expected waiting time until the next offer when a unit has dispatchable battery level $e>\theta$. \\
$h_{(e,j)(e',j')}$ & Sojourn time associated with the transition from state $(e,j)$ to state $(e',j')$. \\
$\bar t^\mathrm{s}, \bar t^\mathrm{r}, \bar t^\mathrm{w}$ 
& Long-run average travel-and-service, return-and-recharging, and waiting-time components, respectively. \\
$P^e(k)$ & Probability that exactly $k$ units are unavailable in the $M/G/N/N$ approximation for energy level $e$. \\
$\varphi^e$ & Offered load recovered by the inverse Erlang method. \\
$C^e$ & Carried load for the approximate $M/G/N/N$ loss system. \\

\addlinespace[4pt]
\multicolumn{2}{@{}l}{\textbf{Decision variables}}\\
\addlinespace[2pt]
$x_{ia}$ & Binary variable equal to 1 if station location $i$ is opened and assigned to zone $a$. \\
$d_{ja}$ & Binary variable equal to 1 if demand node $j$ is assigned to zone $a$. \\
$\bm x^a$ & Station-location vector for zone $a$, $\bm x^a=(x_{ia})_{i\in[M]}$. \\
$\bm d^a$ & Demand-assignment vector for zone $a$, $\bm d^a=(d_{ja})_{j\in[J]}$. \\
$p^a_{ije}$ & Dispatch probability for type-$(j,e)$ demand to station $i$ within zone $a$, implicitly determined by the Energy-Hypercube Iteration. \\
$v_{jj'a}$ & Auxiliary flow from node $j$ to node $j'$ in zone $a$ for contiguity constraints. \\
$h_{ja}$ & Binary variable indicating whether node $j$ is selected as the sink node of zone $a$. \\

$z_{S,I_S}, \bar z_{S,I_S}$ & Decision variable of \textbf{RMP} and its value in the current \textbf{RMP} solution, respectively. \\
$z_{S,n}$ & Aggregated column-selection variable for zone $S$ with $n$ stations. \\
$\bm\pi, \bar{\bm\pi}$ & Dual vector of the restricted master problem and its optimal value in the current iteration, respectively. \\
$\bm d=(d_j)_{j=1}^J$ & Binary zone-membership vector in the pricing problem; $d_j=1$ if node $j$ belongs to the candidate zone. \\
$\bm x=(x_j)_{j=1}^J$ & Binary station-location vector in the pricing problem; $x_j=1$ if a station is opened at node $j$. \\
$\bm d^S, \bm x^{I_S}$ & Incidence vectors of zone $S$ and station set $I_S$, respectively. \\
$y_{ije}$ & Binary assignment variable in the pricing problem, indicating whether demand class $(j,e)$ is assigned to station $i$. \\

\addlinespace[4pt]
\multicolumn{2}{@{}l}{\textbf{Branch-Price-and-Evaluation quantities}}\\
\addlinespace[2pt]

$r_{S,I_S}, \hat r_{S,I_S}, \tilde r_{S,I_S}$ & Exact profit-rate coefficient, surrogate coefficient, and current coefficient of column $(S,I_S)$, respectively. \\
$\hat r^{\mathrm{PM}}_{S,I_S}, \hat r^{\mathrm{UA}}_{S,I_S}$ & $p$-median and uniform-acceptance surrogate coefficients for configuration $(S,I_S)$, respectively. \\
$r_{S,n}$ & Best exact coefficient among configurations with zone $S$ and $n$ stations, $r_{S,n}=\max_{I_S\in\mathcal I(S,n)} r_{S,I_S}$. \\
$\eta_{S,I_S}$ & Zone-level uniform acceptance factor used in the uniform-acceptance surrogate. \\

\addlinespace[4pt]
\multicolumn{2}{@{}l}{\textbf{Numerical experiment quantities}}\\
\addlinespace[2pt]
$R^\star$ & Optimal objective value or profit rate reported in computational comparisons. \\
$T$ & Computational time in seconds. \\
$\Delta R, \Delta T$ & Relative objective-value gap and relative computational-time difference between two methods, respectively. \\
$\alpha$ & Demand scaling factor. \\
$R^{\mathrm{sim}}(x;\alpha)$ & Simulation-evaluated long-run profit rate of decision $x$ under load multiplier $\alpha$. \\
$\widehat R^{\,m}(x;\alpha)$ & Objective value predicted by model $m$ for decision $x$ under load multiplier $\alpha$. \\
$x^m(\alpha)$ & Decision selected by model $m$ under load multiplier $\alpha$. \\
$u^{\mathrm{model}}_{je}, u^{\mathrm{sim}}_{je}$ & Model-predicted and simulation-realized fulfillment probabilities for type-$(j,e)$ demand, respectively. \\
$R(\alpha,A)$ & System-wide profit rate under load multiplier $\alpha$ and zoning intensity $A$. \\

\end{longtable}
\endgroup

\section{Proofs}\label{sec:Proofs}

\subsection{Proof of Lemma~\ref{lemma:irreducibility}}
\label{sec:proof_lemma_irreducibility}

For a fixed unit $i$, let
\[
\mathcal S_i
=
\{(E,b_i)\}
\cup
\{(e,j): e=1,\dots,E-1,\ j=1,\dots,J\}
\]
denote the state space of the embedded chain. This state space is finite.

We next show that the chain is irreducible. For any dispatchable state with $e>\theta$ and any demand node $j'$, Equation~\eqref{eq:Markov_transition} gives
\[
\mathbf P_{(e,j)(e-1,j')}
=
\frac{O_{ij'1}}{\sum_{\epsilon=1}^{e-1} O_{i\epsilon}}
>0.
\]
Hence, from any dispatchable state, the chain can decrease the battery level by one and move to any desired location with positive probability.

We first show that every state can reach the recharging-completion state $(E,b_i)$. If $e\le \theta$, then by Equation~\eqref{eq:Markov_transition},
\[
\mathbf P_{(e,j)(E,b_i)}=1.
\]
If $e>\theta$, choose an arbitrary sequence of demand nodes $j_1,\dots,j_{e-\theta}$. Then the path
\[
(e,j)\to (e-1,j_1)\to \cdots \to (\theta,j_{e-\theta})\to (E,b_i)
\]
has positive probability: each of the first $e-\theta$ transitions has positive probability by the display above, and the last transition occurs with probability one because the battery level is at or below the threshold $\theta$. Therefore, for every state $(e,j)\in\mathcal S_i$, there exists an integer $m\ge 1$ such that
\[
\mathbf P^m_{(e,j)(E,b_i)} > 0.
\]

We now show that $(E,b_i)$ can reach every state in $\mathcal S_i$. Fix any target state $(e',j')\in\mathcal S_i$. If $(e',j')=(E,b_i)$, there is nothing to prove. Otherwise, let $m=E-e'$. Choose arbitrary demand nodes $j_1,\dots,j_{m-1}$. Then the path
\[
(E,b_i)\to (E-1,j_1)\to \cdots \to (e'+1,j_{m-1})\to (e',j')
\]
has positive probability, since each step corresponds to a one-unit energy demand and therefore has strictly positive transition probability. Hence there exists an integer $n\ge 1$ such that
\[
\mathbf P^n_{(E,b_i)(e',j')} > 0.
\]

Combining the two parts, for any two states $x,y\in\mathcal S_i$, there exist integers $m,n\ge 1$ such that
\[
\mathbf P^{m+n}_{xy}
\ge
\mathbf P^m_{x,(E,b_i)}\,
\mathbf P^n_{(E,b_i),y}
>0.
\]
Thus every state communicates with every other state, and the embedded chain is irreducible on the finite state space $\mathcal S_i$.

It follows from standard Markov chain theory that the chain admits a unique stationary distribution $\pi_i$. Moreover, irreducibility on a finite state space implies
\[
\pi_{(e,j)} > 0,\qquad \forall (e,j)\in\mathcal S_i.
\]

\subsection{Proof of Proposition~\ref{prop:semi_markov}}
\label{sec:proof_prop_semi_markov}

For each realized transition $X_n^i=x\to X_{n+1}^i=y$, the next state is sampled according to the embedded-chain transition matrix $\mathbf P_i$, and the associated sojourn time is deterministically equal to $h_{xy}$. Hence, for any states $x$ and $y$,
\[
\mathbb P\bigl(X_{n+1}^i=y,\ T_{n+1}^i-T_n^i\le t\mid X_n^i=x\bigr)
=\mathbf 1_{\{t\ge h_{xy}\}}\,\mathbf P_{xy}.
\]
Therefore, $\{X^i(t)\}_{t\ge 0}$ is a semi-Markov process with embedded chain $\mathbf P_i$ and kernel stated in Proposition~\ref{prop:semi_markov}.

\subsection{Proof of Theorem~\ref{theorem:fixed_point}}
\label{sec:proof_theorem_fixed_point}

Let
\[
x
=
\bigl(\{\rho_i\}_{i=1}^N,\{p_i^e\}_{i=1,\dots,N;\ e=\theta+1,\dots,E}\bigr)
\in \mathcal{K},
\]
and let $T:\mathcal{K}\to\mathcal{K}$ denote the one-step update mapping induced by Algorithm~\ref{algo:iterative_fixed_point}. That is, given $x$, we first compute $\rho_i^{\le e}$, then obtain the inverse-Erlang correction terms, the offer rates, the transition matrices of the embedded Markov chains, and finally the updated quantities
\[
T(x)
=
\bigl(\{\widetilde{\rho}_i\}_{i=1}^N,\{\widetilde{p}_i^e\}_{i=1,\dots,N;\ e=\theta+1,\dots,E}\bigr)
\]
through Equations~\eqref{eq:offer_rate}, \eqref{eq:Markov_transition}, \eqref{eq:rho_time_limit}--\eqref{eq:pie}, and \eqref{eq:loads}--\eqref{eq:r^e}. We prove that $T$ is a continuous self-map on a nonempty compact convex set, and then apply Brouwer's fixed-point theorem.

We proceed in four steps.

\paragraph{Step 1. $\mathcal{K}$ is nonempty, compact, and convex.}
By construction, each interval $[\underline{\rho},\overline{\rho}]$ is compact and convex. For each unit $i$, the set
\[
\Delta_i(\underline{p})
=
\left\{
(p_i^{\theta+1},\dots,p_i^E)\in[\underline{p},1]^{E-\theta}:
\sum_{e=\theta+1}^E p_i^e=1
\right\}
\]
is the intersection of a closed hypercube and an affine hyperplane, and is therefore compact and convex. It is nonempty because $\underline{p}<1/(E-\theta)$. Hence
\[
\mathcal{K}
=
[\underline{\rho},\overline{\rho}]^N
\times
\prod_{i=1}^N \Delta_i(\underline{p})
\]
is also nonempty, compact, and convex.

\paragraph{Step 2. The one-step update mapping $T$ is well defined on $\mathcal{K}$.}
Fix any $x\in\mathcal{K}$. For each unit $i$ and energy threshold $e$, define
\[
\rho_i^{\le e}
=
\rho_i+(1-\rho_i)p_i^{\le e},
\]
where $p_i^{\le e}=1-p_i^{>e}$ and $p_i^{>e}=\sum_{m=e+1}^E p_i^m$. Since $x\in\mathcal{K}$, we have
\[
\underline{\rho}\le \rho_i\le \overline{\rho}
\qquad\text{and}\qquad
p_i^m\ge \underline{p},\quad m=\theta+1,\dots,E.
\]
Therefore, for every $e=\theta+1,\dots,E-1$,
\[
p_i^{>e}
=
\sum_{m=e+1}^E p_i^m
\ge \underline{p},
\]
and hence
\[
\underline{\rho}
\le
\rho_i^{\le e}
=
1-(1-\rho_i)p_i^{>e}
\le
1-(1-\overline{\rho})\underline{p}
<1.
\]
Thus all effective unavailability probabilities are bounded away from both $0$ and $1$ on $\mathcal{K}$.

Next consider the inverse-Erlang step. Since
\[
r^e=\frac{1}{N}\sum_{i=1}^N \rho_i^{\le e},
\]
the above bounds imply that $r^e$ ranges over a compact subinterval of $(0,1)$. For each $e$, the carried-load equation
\[
C^e=\varphi^e(1-P^e(N))
\]
admits a unique solution $\varphi^e$, because the carried load of an $M/G/N/N$ loss system is a strictly increasing function of the offered load. Hence $\varphi^e$, $P^e(k)$, and $Q(N,r^e,k)$ are all well defined.

We now verify the condition of Lemma~\ref{lemma:irreducibility}. Since each preference list is complete, every unit $i$ appears in the preference list of every demand node $j$ at some rank $k$. By assumption, $\lambda_{j1}>0$ for every $j$. Moreover, for every unit ranked ahead of $i$, the corresponding effective unavailability probability is at least $\underline{\rho}>0$. Since $Q(N,r^1,k-1)>0$, Equation~\eqref{eq:offer_rate} yields
\[
O_{ij1}>0,
\qquad \forall i,j.
\]
Hence, for every unit $i$,
\[
O_{i1}=\sum_{j=1}^J O_{ij1}>0.
\]
Therefore, for every battery level $e>\theta$,
\[
\sum_{\epsilon=1}^{e-1} O_{i\epsilon}\ge O_{i1}>0,
\]
so the transition probabilities in Equation~\eqref{eq:Markov_transition} are well defined. Moreover, the condition of Lemma~\ref{lemma:irreducibility} is satisfied for each unit $i$. It follows that the embedded chain is irreducible on a finite state space and therefore admits a unique stationary distribution.

Finally, Equations~\eqref{eq:rho_time_limit}--\eqref{eq:pie} define $\widetilde{\rho}_i$ and $\widetilde{p}_i^e$ from the stationary distributions and state-dependent sojourn times. Since all travel, service, and recharging times are finite and strictly positive, these quantities are well defined. Hence the mapping $T$ is well defined on $\mathcal{K}$.

\paragraph{Step 3. The mapping $T$ is continuous on $\mathcal{K}$.}
We verify continuity componentwise.

First, $\rho_i^{\le e}$ is an affine function of $(\rho_i,p_i^e)$ and is therefore continuous on $\mathcal{K}$. Consequently, $C^e$ and $r^e$ are continuous. Since the inverse-Erlang equation has a unique solution $\varphi^e$ and the carried-load function is strictly increasing in $\varphi^e$, the implicit function theorem implies that $\varphi^e$ is continuous in $C^e$. Therefore, $P^e(k)$ and $Q(N,r^e,k)$ are also continuous.

Second, by Equation~\eqref{eq:offer_rate}, the offer rates $O_{ije}$ are continuous functions of the effective unavailability probabilities. Since the denominators in Equation~\eqref{eq:Markov_transition} are strictly positive on $\mathcal{K}$, the entries of the transition matrix $\mathbf{P}_i$ are continuous functions of $x$.

Third, by Step 2 and Lemma~\ref{lemma:irreducibility}, for each unit $i$, the embedded chain is finite-state and irreducible for every $x\in\mathcal{K}$. Its stationary distribution $\pi_i$ is therefore unique and depends continuously on the transition matrix $\mathbf{P}_i$. Hence $\pi_i$ is continuous in $x$.

Finally, the updated quantities $\widetilde{\rho}_i$ and $\widetilde{p}_i^e$ are obtained from $\pi_i$ and the state-dependent sojourn times through Equations~\eqref{eq:rho_time_limit}--\eqref{eq:pie}. These expressions involve only finitely many additions, multiplications, and divisions by strictly positive denominators, and are therefore continuous. Thus $T$ is continuous on $\mathcal{K}$.

\paragraph{Step 4. Application of Brouwer's fixed-point theorem.}
By assumption (iv), the mapping $T$ satisfies
\[
T(\mathcal{K})\subseteq \mathcal{K}.
\]
By Steps 1 and 3, $\mathcal{K}$ is a nonempty compact convex subset of a finite-dimensional Euclidean space and $T:\mathcal{K}\to\mathcal{K}$ is continuous. Therefore, by Brouwer's fixed-point theorem, there exists
\[
x^* \in \mathcal{K}
\]
such that
\[
T(x^*)=x^*.
\]
That is, Algorithm~\ref{algo:iterative_fixed_point} admits at least one fixed point in $\mathcal{K}$.
\qed

\subsection{Proof of Proposition~\ref{prop:p_median_bound}}
\label{sec:proof_prop_p_median_bound}

Fix a feasible column $(S,I_S)$. Since only open stations can serve demand, we have
\[
p_{ije}=0,\qquad \forall i\notin I_S,\ j\in S,\ e\in\mathcal E.
\]
Hence, by definition,
\[
r_{S,I_S}
=
\sum_{i\in I_S}\sum_{j\in S}\sum_{e\in\mathcal E}
r_{ije}\lambda_{je}p_{ije}
=
\sum_{j\in S}\sum_{e\in\mathcal E}\lambda_{je}
\sum_{i\in I_S} r_{ije}p_{ije}.
\]

For each $(j,e)$, because $r_{ije}\ge 0$ for all $i\in I_S$, and because
\[
p_{ije}\ge 0,
\qquad
\sum_{i\in I_S} p_{ije}\le 1,
\]
we obtain
\[
\sum_{i\in I_S} r_{ije}p_{ije}
\le
\Bigl(\max_{i\in I_S} r_{ije}\Bigr)\sum_{i\in I_S} p_{ije}
\le
\max_{i\in I_S} r_{ije}.
\]
Therefore,
\[
r_{S,I_S}
\le
\sum_{j\in S}\sum_{e\in\mathcal E}\lambda_{je}\max_{i\in I_S} r_{ije}
=
\hat r^{\mathrm{PM}}_{S,I_S}.
\]

It remains to show that $\hat r^{\mathrm{PM}}_{S,I_S}$ is the optimal value of the fixed-set assignment problem. With the open-station set fixed at $I_S$, the problem decomposes across demand classes $(j,e)$. For each $(j,e)$, the constraints
\[
\sum_{i\in I_S} y_{ije}=1,
\qquad
y_{ije}\in\{0,1\},
\]
imply that exactly one station in $I_S$ is selected. Hence the optimal assignment chooses a station attaining the largest reward coefficient, so that
\[
\max_{\substack{y_{ije}\in\{0,1\}\\ \sum_{i\in I_S}y_{ije}=1}}
\sum_{i\in I_S} r_{ije}y_{ije}
=
\max_{i\in I_S} r_{ije}.
\]
Multiplying by $\lambda_{je}$ and summing over all $j\in S$ and $e\in\mathcal E$ gives
\[
\max\left\{
\sum_{i\in I_S}\sum_{j\in S}\sum_{e\in\mathcal E}\lambda_{je}r_{ije}y_{ije}
:
\sum_{i\in I_S} y_{ije}=1,\ 
y_{ije}\in\{0,1\}
\right\}
=
\hat r^{\mathrm{PM}}_{S,I_S}.
\]
This proves the proposition.

\subsection{Proof of Theorem~\ref{theorem:BPE_convergency}}
\label{sec:proof_theorem_BPE_convergency}

Fix an arbitrary branch-and-bound node $n$, and let $\mathbb{S}(n)$ denote the set of all
columns $(S,I_S)$ that satisfy the branching decisions associated with node $n$. By
assumption, $\mathbb{S}(n)$ is finite. The LP relaxation at node $n$ is exactly the master
problem restricted to the columns in $\mathbb{S}(n)$.

For each node-feasible column $(S,I_S)\in \mathbb{S}(n)$, let $r_{S,I_S}$ denote its exact
coefficient and let $\tilde r_{S,I_S}$ denote the coefficient currently used by
Algorithm~\ref{algo:CGE}. By assumption, we have $\tilde r_{S,I_S}\ge r_{S,I_S}$ for every
$(S,I_S)\in \mathbb{S}(n)$. At node $n$, Algorithm~\ref{algo:CGE} alternates between two
steps: (i) solving the restricted master problem with the current coefficients
$\tilde r_{S,I_S}$ and generating improving columns through pricing, and (ii) exactly
evaluating every active column and replacing its surrogate coefficient by its exact value.
Therefore, at a fixed node, Algorithm~\ref{algo:CGE} is a direct specialization of the generic
coefficient-update scheme analyzed in Theorem~2 of
\citet{ZhangEtAl2023ScenarioReduction}. Since the number of node-feasible columns is finite,
Theorem~2 of \citet{ZhangEtAl2023ScenarioReduction} implies that
Algorithm~\ref{algo:CGE} finitely converges to the optimal solution of the LP relaxation at
node $n$.

Hence, every processed node in Algorithm~\ref{algo:BPE} is solved exactly to its node-optimal
LP-relaxation value. Since \textbf{MP} is a maximization problem, this value is a valid upper
bound on all integer solutions feasible to that node. Moreover, the branching scheme partitions
the remaining feasible integer solutions into disjoint subsets without excluding any feasible
optimal solution. Because the master problem admits only finitely many feasible columns, the
number of feasible integer solutions is finite, and thus the branch-and-bound tree is finite as
well. Standard branch-and-bound arguments then imply that, once each node is solved exactly to
its LP-relaxation optimum, Algorithm~\ref{algo:BPE} terminates finitely and returns an optimal
integer solution of \textbf{MP}.

\subsection{Aggregation Equivalence of \textbf{MP} and \textbf{MP$'$}}
\label{sec:proof_prop_MP'_equiv}

We begin with the integer program 

{\SingleSpacedXI
\begin{subequations}\label{MP'}
\begin{align}
\textbf{MP$'$:}\quad\max\ & \sum_{(S,n)\in\mathbb{S}'} r_{S,n}\, z_{S,n} \notag\\
\text{s.t.}\
& \sum_{\substack{(S,n)\in\mathbb{S}'\\ j\in S}} z_{S,n} = 1, &&\forall j,  \label{MP':cover}\\
& \sum_{(S,n)\in\mathbb{S}'} n\, z_{S,n} = N,
&&  \label{MP':stations}\\
& \sum_{(S,n)\in\mathbb{S}'} z_{S,n} = A,
&&  \label{MP':zones}\\
& z_{S,n}\in\{0,1\}, &&\forall (S,n)\in\mathbb{S}'. \label{MP':binary}
\end{align}
\end{subequations}
}
where $\mathbb{S}'=\{(S,n):\exists (S,I_S)\in\mathbb{S}, |I_S|=n\}$. 

\begin{lemma}\label{lem:MP'_equiv}
For each $(S,n)\in\mathbb S'$, let
\[
\mathcal I(S,n):=\{I_S\subseteq S:(S,I_S)\in\mathbb S,\ |I_S|=n\},
\qquad
r_{S,n}:=\max_{I_S\in\mathcal I(S,n)} r_{S,I_S}.
\]
Under the aggregation
\[
z_{S,n}:=\sum_{I_S\in\mathcal I(S,n)} z_{S,I_S},
\qquad \forall (S,n)\in\mathbb S',
\]
\textbf{MP$'$} is an exact aggregation of \textbf{MP}. In particular, \textbf{MP} and \textbf{MP$'$} are equivalent and have the same optimal objective value.
\end{lemma}

We prove the claim by showing a two-way correspondence between feasible solutions of \textbf{MP} and \textbf{MP$'$}.

First, let $\bm z=\{z_{S,I_S}\}_{(S,I_S)\in\mathbb S}$ be any feasible solution of \textbf{MP}, and define
\[
z_{S,n}:=\sum_{I_S\in\mathcal I(S,n)} z_{S,I_S},
\qquad \forall (S,n)\in\mathbb S'.
\]
Because all columns aggregated into $z_{S,n}$ share the same zone $S$ and the same cardinality $n$, the coverage, station-budget, and zone-count constraints are preserved under this transformation. Moreover, two distinct selected columns in \textbf{MP} cannot share the same zone $S$, since otherwise every demand node in $S$ would be covered more than once. Hence $z_{S,n}\in\{0,1\}$ for all $(S,n)\in\mathbb S'$, so $\{z_{S,n}\}$ is feasible for \textbf{MP$'$}.

By definition of $r_{S,n}$, we have $r_{S,n}\ge r_{S,I_S}$ for every $I_S\in\mathcal I(S,n)$. Therefore,
\[
\sum_{(S,n)\in\mathbb S'} r_{S,n} z_{S,n}
=
\sum_{(S,n)\in\mathbb S'} r_{S,n}\sum_{I_S\in\mathcal I(S,n)} z_{S,I_S}
\ge
\sum_{(S,n)\in\mathbb S'} \sum_{I_S\in\mathcal I(S,n)} r_{S,I_S} z_{S,I_S}
=
\sum_{(S,I_S)\in\mathbb S} r_{S,I_S} z_{S,I_S}.
\]
Hence every feasible solution of \textbf{MP} induces a feasible solution of \textbf{MP$'$} with objective value no smaller, implying
\[
\operatorname{OPT}(\textbf{MP})\le \operatorname{OPT}(\textbf{MP$'$}).
\]

Conversely, let $\{z_{S,n}\}_{(S,n)\in\mathbb S'}$ be any feasible solution of \textbf{MP'}. For each active pair $(S,n)$ with $z_{S,n}=1$, choose
\[
I_S^*(S,n)\in \arg\max_{I_S\in\mathcal I(S,n)} r_{S,I_S}.
\]
Construct a solution of \textbf{MP} by setting
\[
z_{S,I_S}=
\begin{cases}
1, & \text{if } z_{S,n}=1 \text{ and } I_S=I_S^*(S,n),\\
0, & \text{otherwise.}
\end{cases}
\]
Since each active variable $z_{S,n}$ is replaced by exactly one column with the same zone $S$ and the same cardinality $n$, the resulting solution satisfies the coverage, station-budget, and zone-count constraints of \textbf{MP}. Its objective value is
\[
\sum_{(S,I_S)\in\mathbb S} r_{S,I_S} z_{S,I_S}
=
\sum_{(S,n)\in\mathbb S'} r_{S,I_S^*(S,n)} z_{S,n}
=
\sum_{(S,n)\in\mathbb S'} r_{S,n} z_{S,n}.
\]
Thus every feasible solution of \textbf{MP$'$} can be lifted to a feasible solution of \textbf{MP} with the same objective value, implying
\[
\operatorname{OPT}(\textbf{MP$'$})\le \operatorname{OPT}(\textbf{MP}).
\]

Combining the two inequalities yields
\[
\operatorname{OPT}(\textbf{MP})=\operatorname{OPT}(\textbf{MP$'$}),
\]
which proves that \textbf{MP'} is an exact aggregation of \textbf{MP}.

The same aggregation argument applies to the corresponding restricted master formulations: \textbf{RMP$'$} is obtained from \textbf{RMP} by replacing columns indexed by $(S,I_S)$ with aggregated columns indexed by $(S,n)$.

\section{Full MILP Formulation of the P-Median Surrogate Pricing Problem}\label{sec:app_spp}

This appendix gives the full MILP formulation of the $p$-median surrogate
pricing problem used in Section~\ref{sec:pm-surrogate}. Given an optimal dual
vector $\bar{\bm\pi}$ of the current \textbf{RMP}, let $d_j$ indicate whether
node $j$ is selected in the candidate zone, let $x_j$ indicate whether a
station is opened at node $j$, and let $y_{ije}$ indicate whether demand class
$(j,e)$ is assigned to station $i$. The full formulation is

{\SingleSpacedXI
\begin{subequations}\label{SPP-linear}
\begin{align}
\textbf{SPP:}\quad \max\ &
\sum_{i\in[J]}\sum_{j\in[J]}\sum_{e\in\mathcal E}
\lambda_{je} r_{ije} y_{ije}
-\sum_{j\in[J]} \bar\pi_j d_j
-\bar\pi_{J+1}\sum_{i\in[J]} x_i
-\bar\pi_{J+2} \notag\\
\textup{s.t.}\quad
& \sum_{i\in[J]} y_{ije}=d_j,
&& \forall j\in[J],\ e\in\mathcal E, \label{SPP:assign}\\
& y_{ije}\le x_i,
&& \forall i\in[J],\ j\in[J],\ e\in\mathcal E, \label{SPP:assign_open}\\
& 1\le \sum_{j\in[J]} x_j \le N-A+1, \label{SPP:station_num}\\
& x_j \le d_j,
&& \forall j\in[J], \label{SPP:x_implies_d}\\
& \sum_{j':(j,j')\in\mathcal{A}} v_{jj'}
  - \sum_{j'':(j'',j)\in\mathcal{A}} v_{j''j}
    \ge d_j-(J-A+1)h_j,
&& \forall j\in[J], \label{SPP:cont1}\\
& v_{jj'}+v_{j'j}\le (J-A)d_j,
&& \forall (j,j')\in\mathcal{A}, \label{SPP:cont2}\\
& \sum_{j\in [J]} h_j = 1, \label{SPP:cont3}\\
& h_j\le d_j,
&& \forall j\in[J], \label{SPP:cont4}\\
& y_{ije}\in\{0,1\},
&& \forall i\in[J],\ j\in[J],\ e\in\mathcal E, \label{SPP:y_binary}\\
& x_j,h_j,d_j\in\{0,1\},
&& \forall j\in[J], \label{SPP:binary}\\
& v_{jj'}\ge 0,
&& \forall (j,j')\in\mathcal{A}. \label{SPP:nonneg}
\end{align}
\end{subequations}
}
Constraints~\eqref{SPP:assign}--\eqref{SPP:assign_open} encode the assignment
structure induced by the $p$-median surrogate: if $d_j=1$, each demand class
$(j,e)$ is assigned to exactly one station, and if $d_j=0$, no assignment is
allowed; assignments can only be made to open stations. Constraints
\eqref{SPP:station_num} and \eqref{SPP:x_implies_d} enforce the admissible
station count for a single column and require opened stations to lie inside
the selected zone. Constraints~\eqref{SPP:cont1}--\eqref{SPP:cont4} are the
same flow-based contiguity constraints as in \textbf{LZP}, adapted to a single
candidate zone. Constraints~\eqref{SPP:y_binary}--\eqref{SPP:nonneg} define
the variable domains.

\section{Derivation of Uniform-Acceptance Surrogate and Approximate Pricing Problem.}\label{sec:app_ua_surrogate}

The $p$-median surrogate is valid but can be loose when service capacity is tight because it implicitly treats all demand as fully serviceable. To obtain a sharper surrogate for pricing, we reinterpret $\hat r^{\mathrm{PM}}_{S,I_S}$ as a full-acceptance reward and then scale it by a zone-level acceptance factor that captures blocking in an aggregate way. This yields a surrogate that remains compatible with MILP pricing while better reflecting how a higher offered load reduces the effective reward through blocking. In practice, it is substantially tighter than the $p$-median surrogate and is the pricing model used in most of our numerical experiments.

Recall from Proposition~\ref{prop:p_median_bound} that
$
\hat r^{\mathrm{PM}}_{S,I_S}
=\sum_{j\in S}\sum_{e\in\mathcal E}\lambda_{je}\max_{i\in I_S}r_{ije}.
$
This quantity can be interpreted as the reward obtained under full acceptance, i.e., when every demand is served by the open station in $I_S$ that yields the largest reward. Since $r_{ije}=r_e-c_{ij}$, we may rewrite it as
$
\hat r^{\mathrm{PM}}_{S,I_S}
=\sum_{j\in S}\Bigl(\sum_{e\in\mathcal E}\lambda_{je}r_e\Bigr)
-\sum_{j\in S}\Bigl(\sum_{e\in\mathcal E}\lambda_{je}\Bigr)c_j^{\min},$ where $
c_j^{\min}:=\min_{i\in I_S}c_{ij}.
$
Let
$
\beta_j:=\sum_{e\in\mathcal E}\lambda_{je}r_e,\;
\gamma_j:=\sum_{e\in\mathcal E}\lambda_{je}.
$
In the pricing MILP, the nearest open station for each $j$ is identified by binary variables $z_{ij}^{\min}$:
\par\vspace{-0.4\baselineskip}
{\SingleSpacedXI
\begin{subequations}\label{eq:uas_zmin}
\begin{align}
\sum_{i=1}^J z_{ij}^{\min} &= d_j, &&\forall j\in[J],\\
z_{ij}^{\min} &\le x_i, &&\forall i,j\in[J],\\
c_j^{\min} &= \sum_{i=1}^J c_{ij}z_{ij}^{\min}, &&\forall j\in[J].
\end{align}
\end{subequations}
}
Hence, the full-acceptance reward can be written as
\[
\hat r^{\mathrm{PM}}_{S,I_S}
=\sum_{j=1}^J \beta_j d_j
-\sum_{i=1}^J\sum_{j=1}^J \gamma_j c_{ij} z_{ij}^{\min}.
\]

To account for blocking, we next introduce a uniform acceptance factor. Let $n:=|I_S|=\sum_i x_i$, $\lambda_j:=\sum_{e\in\mathcal E}\lambda_{je}$, and $\lambda_{\mathrm{tot}}:=\sum_j \lambda_j d_j$. We approximate the pooled offered load by the conservative expression
\[
a^{\mathrm{lb}}
:=
\sum_{j\in[J]}\sum_{e\in\mathcal E}\lambda_{je}d_j\,\tau^s_e
+\sum_{j\in[J]} \lambda_j d_j\,\tau_j^t
+\frac{\tau^r}{\Delta^E}\sum_{j\in[J]}\sum_{e\in\mathcal E}\lambda_{je}d_j\,e,
\]
where $\Delta^E:=\max\{1,(E-1)-\theta\}$, $\tau^r$ is the average recharge time, and $\tau_j^t:=\min_{i:x_i=1}\tau_{ij}^t$ is the travel time from node $j$ to its nearest open station. The nearest-station travel time $\tau_j^t$ can be linearized as follows:
\par\vspace{-0.4\baselineskip}
{\SingleSpacedXI
\begin{subequations}\label{eq:uas_travel}
\begin{align}
\sum_{i=1}^J z^t_{ij} &= d_j, &&\forall j\in[J],\\
z^t_{ij} &\le x_i, &&\forall i,j\in[J],\\
\tau_j^t &= \sum_{i=1}^J \tau_{ij}^t z^t_{ij}, &&\forall j\in[J].
\end{align}
\end{subequations}
}

The quantity $a^{\mathrm{lb}}$ has a conservative interpretation: the first term captures the service-load contribution generated by the demand rates and service times, the nearest-station travel term is no larger than the actual dispatch travel, and the third term approximates recharge load as the fixed time per recharge multiplied by the average number of recharge activities per unit time, obtained by converting aggregate energy consumption into equivalent recharge counts. 

Given offered load $a$ and $n$ identical servers, let
$
B(n,a)=\frac{a^n/n!}{\sum_{k=0}^n a^k/k!}
$
denote the Erlang-B blocking probability. For each feasible $n$, we construct a piecewise-linear lower approximation $B_n^{\mathrm{lb}}(\cdot)\le B(n,\cdot)$ and define the uniform acceptance factor by
$
\eta_{S,I_S}:=1-B_n^{\mathrm{lb}}(a^{\mathrm{lb}}).
$
Since blocking is increasing in the offered load, $\eta_{S,I_S}$ is an upper estimate of the common acceptance rate used in pricing. We then define the uniform-acceptance (UA) surrogate as
\[
\hat r^{\mathrm{UA}}_{S,I_S}
:=\eta_{S,I_S}\hat r^{\mathrm{PM}}_{S,I_S}
=(1-B_n^{\mathrm{lb}}(a^{\mathrm{lb}}))\hat r^{\mathrm{PM}}_{S,I_S}.
\]

Because $\hat r^{\mathrm{PM}}_{S,I_S}$ is linear in the binary variables once \eqref{eq:uas_zmin} is imposed, the product above can be linearized exactly. Introduce auxiliary variables
$
y_j^d:=\eta_{S,I_S}d_j, \;
y_{ij}^{\min}:=\eta_{S,I_S}z_{ij}^{\min},
$
with $\eta_{S,I_S}\in[0,1]$. Their exact linearization is

{\SingleSpacedXI
\begin{equation}\label{eq:uas_exactlin}
\left\{
\begin{aligned}
&0 \le y_j^d \le \eta_{S,I_S},
&&y_j^d \le d_j,
&&y_j^d \ge \eta_{S,I_S}-(1-d_j),
&& \forall j\in[J],\\
&0 \le y_{ij}^{\min} \le \eta_{S,I_S},
&&y_{ij}^{\min} \le z_{ij}^{\min},
&&y_{ij}^{\min} \ge \eta_{S,I_S}-(1-z_{ij}^{\min}),
&& \forall i,j\in[J].
\end{aligned}
\right.
\end{equation}
}
The surrogate coefficient can therefore be written as
\begin{equation}\label{eq:uas_reward}
\hat r^{\mathrm{UA}}_{S,I_S}
=\sum_{j=1}^J \beta_j y_j^d
-\sum_{i=1}^J\sum_{j=1}^J \gamma_j c_{ij} y_{ij}^{\min}.
\end{equation}

Replacing the pricing objective by \eqref{eq:uas_reward} yields the approximate surrogate pricing problem:

{\SingleSpacedXI
\begin{subequations}\label{eq:SPP_UAS}
\begin{align*}
\textbf{aSPP}:\quad
\max\quad
&\sum_{j=1}^J \beta_j y_j^d
-\sum_{i=1}^J\sum_{j=1}^J \gamma_j c_{ij} y_{ij}^{\min}
-\sum_{j=1}^J \bar\pi_j d_j
-\bar\pi_{J+1}\sum_{i=1}^J x_i
-\bar\pi_{J+2} \\
\text{s.t.}\quad
& \eqref{SPP:station_num}-\eqref{SPP:nonneg},\ \eqref{constr:updated_r},\\
& \eqref{eq:uas_zmin},\ \eqref{eq:uas_travel},\\
& a^{\mathrm{lb}}
=\sum_{j\in[J]}\sum_{e\in\mathcal E}\lambda_{je}d_j\,\tau_e^s
+\sum_{j\in[J]} \lambda_j d_j\,\tau_j^t
+\frac{\tau^r}{\Delta^E}\sum_{j\in[J]}\sum_{e\in\mathcal E}\lambda_{je}d_j\,e,\\
& \eta_{S,I_S}=1-B_n^{\mathrm{lb}}(a^{\mathrm{lb}}),
\qquad n=\sum_i x_i,
\qquad 0\le \eta_{S,I_S}\le 1,\\
& \eqref{eq:uas_exactlin}.
\end{align*}
\end{subequations}
}

The exact BPE framework requires the pricing surrogate to upper-bound the exact column coefficient for every feasible column. Proposition~\ref{prop:p_median_bound} guarantees this property for the $p$-median surrogate. The UA surrogate, by contrast, does not currently admit such a theoretical guarantee. Accordingly, we use it as an empirically motivated surrogate rather than as a theoretically certified upper bound. In our computational study, however, the UA surrogate remains upper-bounding on the tested dataset and is substantially tighter than the $p$-median surrogate, which makes it much more effective for practical pricing.

\begin{table}[t]
\centering
{\SingleSpacedXI
\caption{Comparison of the uniform acceptance surrogate and the $p$-median surrogate on the synthetic Toronto subinstance.}
\label{tab:surrogate_comparison}
}
\vspace{0.5em}
{\SingleSpacedXI
\footnotesize
\begin{tabular}{lccccccc}
\toprule
Surrogate & $\Pr(\hat r \ge r)$ & $\min \frac{\hat r-r}{r}(\%)$ & RMSE & MAE & MAPE (\%) & Pearson corr. & Spearman corr. \\
\midrule
UA & $100.0\%$ & $8.3859$ & $1.2851$ & $1.0596$ & $20.4830$ & $0.9974$ & $0.9960$ \\
p-median & $100.0\%$ & $15.6530$ & $11.4520$ & $10.5223$ & $346.1026$ & $0.3872$ & $0.3833$ \\
\bottomrule
\end{tabular}
}
\vspace{0.7em}
{\SingleSpacedXI
\footnotesize
\noindent\parbox{0.96\linewidth}{
\textit{Notes.} $\Pr(\hat r \ge r)$ is the fraction of tested columns for which the surrogate value is no smaller than the exact value. The minimum percentage gap is computed as $\min\bigl((\hat r-r)/r \times 100\%\bigr)$. Pearson corr.\ and Spearman corr.\ denote the Pearson correlation coefficient and the Spearman rank correlation coefficient between $\hat r$ and $r$, respectively.
}
}
\end{table}

Table~\ref{tab:surrogate_comparison} reports the comparison on 9,908 candidate columns generated from the synthetic Toronto subinstance. Both surrogates are upper-bounding on this dataset, but the UA surrogate is far tighter and much better aligned with the exact coefficient in both magnitude and rank. In particular, the Pearson and Spearman correlations between $\hat r^{\mathrm{UA}}$ and $r$ are close to one, whereas the $p$-median surrogate is much less informative about the relative quality of columns. This empirical evidence explains why \textbf{aSPP} is substantially more effective than \textbf{SPP} in practice. Because the UA surrogate is much tighter, the resulting pricing problem typically converges faster. In addition, since it more closely preserves the ranking induced by the exact coefficients, it is more likely to identify high-quality columns, thereby improving the quality of column generation.

\section{Detailed Branching Rules and Pseudocode for the BPE Framework}
\label{sec:app_bpe_details}

\subsection{Branching Rules in the Exact BPE Framework}
\label{sec:app_branching}

The exact BPE framework uses a two-phase branching rule. We first apply a Ryan--Foster-type branching rule \citep{ryan1981integer} based on pairwise zone-membership relations, because it acts directly on the partition structure and typically yields stronger child-node restrictions. When no effective pairwise branch remains, we fall back on conventional branching on a fractional column-selection variable $z_{S,I_S}$.

For pairwise branching, let $\mathcal{B}^d$ denote the set of active pairwise branching decisions. For each $(j,j')\in\mathcal{B}^d$, the two child-node branches are
\begin{subequations}\label{eq:branching_d}
\begin{align}
 d_j=d_{j'},
 &\qquad \text{for the branch } d_{jj'}=1, \label{constr:branch-d1}\\
 d_j+d_{j'}\le 1,
 &\qquad \text{for the branch } d_{jj'}=0. \label{constr:branch-d0}
\end{align}
\end{subequations}
The first branch forces demand nodes $j$ and $j'$ to appear in the same generated zone, whereas the second forbids them from appearing in the same generated zone.

For column branching, let $\mathcal{B}^z$ denote the set of active column-branching decisions. For each $(S,I_S)\in\mathcal{B}^z$, let
\[
\bm d^S=(d_j^S)_{j=1}^J,\qquad d_j^S:=\mathbf{1}\{j\in S\},
\]
and
\[
\bm x^{I_S}=(x_j^{I_S})_{j=1}^J,\qquad x_j^{I_S}:=\mathbf{1}\{j\in I_S\}
\]
denote the indicator vectors of the branched column. The two child-node branches are
\begin{subequations}\label{eq:branching_z}
\begin{align}
\sum_{j=1}^J d_j d_j^S = 0,\qquad
\sum_{j=1}^J x_j x_j^{I_S} = 0,
&\qquad \text{for the branch } z_{S,I_S}=1, \label{constr:branch-z1}\\
\sum_{j=1}^J \bigl(d_j+d_j^S-2d_jd_j^S\bigr)
+\sum_{j=1}^J \bigl(x_j+x_j^{I_S}-2x_jx_j^{I_S}\bigr)\ge 1,
&\qquad \text{for the branch } z_{S,I_S}=0. \label{constr:branch-z0}
\end{align}
\end{subequations}
The branch $z_{S,I_S}=1$ fixes column $(S,I_S)$ to one and restricts subsequent pricing to the residual instance, so no newly generated column may use demand nodes or station locations already covered by $(S,I_S)$. The branch $z_{S,I_S}=0$ excludes column $(S,I_S)$ from future pricing by requiring every generated column to differ from $(S,I_S)$ in at least one component.

All active branching constraints are imposed directly in the pricing problem at the corresponding node, so that only branch-compatible columns can be generated. Existing master columns that violate the node's branching decisions are removed from the corresponding restricted master problem.

\subsection{Detailed Algorithmic Logic}
\label{sec:app_bpe_logic}

Algorithm~\ref{algo:CGE} gives the node-level column-generation-and-evalution routine. Starting from the current restricted master problem and the branching decisions inherited by a node, the procedure first removes all master columns that are incompatible with the node restrictions. It then alternates between surrogate pricing and exact coefficient evaluation until the node LP relaxation is solved exactly.

Algorithm~\ref{algo:BPE} embeds Algorithm~\ref{algo:CGE} into a branch-and-bound tree. Let $\mathcal{N}^{\mathrm{BPE}}:=\{n_i:i=0,1,\ldots\}$ denote the set of branch-and-bound nodes, and let $\mathcal{Q}\subseteq \mathcal{N}^{\mathrm{BPE}}$ be the queue of active nodes whose LP relaxations have been solved but that have not yet been fathomed. The state of node $n_i$ is represented by
$
n_i(\tilde{\bm r}_i,\tilde{\mathbb S}_i,\mathbb S_i^u,\mathcal{B}_i^d,\mathcal{B}_i^z),
$
where $(\tilde{\bm r}_i,\tilde{\mathbb S}_i,\mathbb S_i^u)$ records the current status of $\textbf{RMP}(\tilde{\bm r}_i,\tilde{\mathbb S}_i)$, and $\mathcal{B}_i^d$ and $\mathcal{B}_i^z$ record the pairwise and column-branching decisions associated with that node. Child nodes inherit the parent state $(\tilde{\bm r}_i,\tilde{\mathbb S}_i,\mathbb S_i^u)$ and append one new branching decision; before reoptimization, all master columns incompatible with the child-node restrictions are removed.

\newpage
\begin{algorithm}[!htbp]
\caption{Column-Generation-and-Evaluation}\label{algo:CGE}
{\footnotesize
\begin{algorithmic}[1]
\State \textbf{Input:} $\tilde{\bm r}, \tilde{\mathbb{S}}, \mathcal{B}^d, \mathcal{B}^z, \mathbb{S}^u$.
\State Remove from $\tilde{\mathbb{S}}$ all columns incompatible with $(\mathcal{B}^d,\mathcal{B}^z)$, and update $\mathbb{S}^u \leftarrow \mathbb{S}^u \cap \tilde{\mathbb{S}}$.
\While{\textbf{true}}
    \While{\textbf{true}}
        \State Solve $\textbf{RMP}(\tilde{\bm r},\tilde{\mathbb{S}})$, obtain the optimal solution $\bar{\bm z}$ and dual variables $\bar{\bm \pi}$.
        \State Add branching constraints \eqref{constr:branch-d1}--\eqref{constr:branch-z0} based on $\mathcal{B}^d,\mathcal{B}^z$, and exclusion constraints \eqref{constr:updated_r} based on $\mathbb{S}^u$, to \textbf{SPP}/\textbf{aSPP}.
        \State Solve \textbf{SPP}/\textbf{aSPP}; let $rc$ be its optimal objective value.
        \If{$rc>0$}
            \State Recover the generated column $(S,I_S)$ and its surrogate coefficient $\hat r_{S,I_S}$.
            \State Add $(S,I_S)$ into $\tilde{\mathbb{S}}$ and set $\tilde r_{S,I_S}\leftarrow \hat r_{S,I_S}$.
        \Else
            \State Break.
        \EndIf
    \EndWhile
    \If{$\mathcal{C}(\bar{\bm z})\subseteq \mathbb{S}^u$}
        \State $v^{\mathrm{opt}} \leftarrow \sum_{(S,I_S)\in\tilde{\mathbb{S}}}\tilde r_{S,I_S}\bar z_{S,I_S}$.
        \Return  $v^{\mathrm{opt}}$,  $\bar{\bm z}$, and the updated $(\tilde{\bm r},\tilde{\mathbb{S}},\mathbb{S}^u,\mathcal{B}^d,\mathcal{B}^z)$.
    \EndIf
    \State For each $(S,I_S)\in\mathcal{C}(\bar{\bm z})\setminus \mathbb{S}^u$, compute its exact coefficient $r_{S,I_S}$ by Algorithm~\ref{algo:iterative_fixed_point}.
    \State Update $\tilde r_{S,I_S}\leftarrow r_{S,I_S}$ for all $(S,I_S)\in\mathcal{C}(\bar{\bm z})\setminus \mathbb{S}^u$.
    \State Update $\mathbb{S}^u\leftarrow \mathbb{S}^u\cup \mathcal{C}(\bar{\bm z})$.
\EndWhile
\end{algorithmic}
}
\end{algorithm}

\newpage

\begin{algorithm}[!htbp]
\caption{Branch-Price-and-Evaluation}\label{algo:BPE}
{\footnotesize
\begin{algorithmic}[1]
\State \textbf{Input:} all problem data of \textbf{MP}, including $\lambda_{je},\tau^t_{jj'}, \tau^s_e, \tau^{rc}_e, r_e, c_{ij}$, the station budget $N$, the zone count $A$, and the graph $\mathcal{G}=(\mathcal{N},\mathcal{A})$.
\State \textbf{Initialize:} choose feasible $\tilde{\bm r}^{\,0}, \tilde{\mathbb{S}}^{\,0}$. Let $\mathcal{Q}=\emptyset$, $\bar n$ be None, $\bar v=-\infty$, and $\bar{\bm z}=\mathbf{0}_{|\mathbb{S}|}$.
\State Create the root node $n_0(\tilde{\bm r}^{\,0}, \tilde{\mathbb{S}}^{\,0}, \emptyset, \emptyset, \emptyset)$.
\State Solve $n_0$ by Algorithm~\ref{algo:CGE}, obtain $v_0^{\mathrm{opt}}$, $\tilde{\bm z}^{*}_0$, and the updated node state $(\tilde{\bm r}_0,\tilde{\mathbb{S}}_0,\mathbb{S}_0^u)$.
\If{$n_0$ is feasible}
    \State Store $(v_0^{\mathrm{opt}},\tilde{\bm z}^{*}_0)$ in $n_0$ and push $n_0$ into $\mathcal{Q}$.
\EndIf
\While{$\mathcal{Q}\neq \emptyset$}
    \State Get a node $n_k\in\mathcal{Q}$ and remove it from $\mathcal{Q}$.
    \State Read from $n_k$ its stored node-optimal value $v_k^{\mathrm{opt}}$ and optimal solution $\tilde{\bm z}^{*}_k$.
    \If{$v_k^{\mathrm{opt}}\le \bar v$}
        \State Prune node $n_k$.
    \ElsIf{$\tilde{\bm z}^{*}_k$ is integral}
        \State $\bar n\leftarrow n_k$, $\bar v\leftarrow v_k^{\mathrm{opt}}$, $\bar{\bm z}\leftarrow \tilde{\bm z}^{*}_k$.
    \Else
        \State Apply the two-phase branching rule to $\tilde{\bm z}^{*}_k$.
        \If{a Ryan--Foster pair $(j,j')$ is selected}
            \State Create two child nodes by appending the decisions $d_{jj'}=1$ and $d_{jj'}=0$, respectively.
        \Else
            \State Select a fractional column $(S,I_S)$ with $0<(\tilde z^{*}_{k})_{S,I_S}<1$.
            \State Create two child nodes by appending the decisions $z_{S,I_S}=1$ and $z_{S,I_S}=0$, respectively.
        \EndIf
        \For{each child node $n_{k_i}$, $i\in\{1,2\}$}
            \State Let $n_{k_i}$ inherit the parent state $(\tilde{\bm r}_k,\tilde{\mathbb{S}}_k,\mathbb{S}_k^u)$ together with the corresponding updated branching decisions.
            \State Solve $n_{k_i}$ by Algorithm~\ref{algo:CGE}, obtain $v_{k_i}^{\mathrm{opt}}$, $\tilde{\bm z}^{*}_{k_i}$, and the updated node state.
            \If{$n_{k_i}$ is feasible and $v_{k_i}^{\mathrm{opt}}>\bar v$}
                \State Store $(v_{k_i}^{\mathrm{opt}},\tilde{\bm z}^{*}_{k_i})$ in $n_{k_i}$ and push $n_{k_i}$ into $\mathcal{Q}$.
            \EndIf
        \EndFor
    \EndIf
\EndWhile
\Return $\bar v,\bar{\bm z}$.
\end{algorithmic}
}
\end{algorithm}

\newpage

\section{Settings of Real Cases}\label{sec:app_params}

The real-case instances are constructed on the Toronto Forward Sortation Area (FSA) layer. An FSA is identified by the first three characters of a Canadian postal code. We use the 2021 Census FSA boundary file from Statistics Canada and select the 36 connected FSAs in the study region; the centroid of each FSA is used as one demand node \citep{statcan2021FSABoundary}. FSA-level population counts from the 2021 Census are used to scale the baseline arrival rates \citep{statcan2021FSAPopulation}. Specifically, for application $c$ and demand node $j$, we set
\[
\Lambda^{(c)}_j
=
\alpha \lambda_c \frac{\mathrm{Pop}_j}{10^6},
\]
where $\alpha$ is the scaling factor, $\lambda_c=0.10$ for the mobile-charging and drone-inspection cases, and $\lambda_c=0.05$ for the autonomous-cleaning case. Here, population is used as a proxy for potential demand intensity.

Let $d_{jj'}$ denote the distance between demand nodes $j$ and $j'$ in kilometers. The travel time is computed as
\[
\tau^{t(c)}_{jj'}
=
60 \frac{d_{jj'}}{v_c},
\]
where $v_c$ is the case-specific travel speed in km/hour. The same rule is used for travel from a demand node back to the base station. For the mobile-charging case, we set $v_c=16$ km/hour. Since movable charger products usually report customer-facing output power rather than detailed autonomous travel specifications, this value is used as a conservative mobility assumption, benchmarked against congested urban traffic in Toronto \citep{tomtomTorontoTraffic2025}. For the drone-inspection case, we set $v_c=36$ km/hour, or 10 m/s, which is consistent with the operating speed used in enterprise-drone endurance specifications \citep{djiMatrice4SeriesSpecs}. For the autonomous-cleaning case, we set $v_c=5$ km/hour, consistent with the operating-speed range of autonomous and robotic sweepers \citep{trombiaFree,puduMT1}.

Table~\ref{tab:real_case_settings} summarizes the main application-specific parameters. In the mobile-charging case, one energy level is 16 kWh and the battery capacity is $E=10$ levels, giving a total capacity of 160 kWh. This capacity is benchmarked against FreeWire's battery-integrated DC fast charger, which reports a 160 kWh battery capacity \citep{freewireBoostCharger150}. A one-level service takes 20 minutes, implying an on-site discharge rate of about 48 kW. This is an advanced but commercially observed setting, consistent with movable or mobile DC chargers whose reported output power is in the 40--50 kW range \citep{siemensHelioxMobile40,lincolnVelion50Mobile,kempowerMovableCharger}.

In the drone-inspection case, one energy level is 10 Wh and $E=10$, giving a total capacity of 100 Wh, which is close to the battery capacity of enterprise inspection UAVs \citep{djiMatrice4SeriesSpecs}. The on-site service stage is interpreted as hovering or low-speed loitering while the drone performs visual, thermal, or asset-inspection tasks. A one-level service takes 10 minutes, so the implied service-stage energy rate is 60 W. This value is of the same order as the average power implied by the battery capacity and endurance of DJI's enterprise inspection UAVs \citep{djiMavic3EnterpriseSpecs,djiMatrice4SeriesSpecs}.

In the autonomous-cleaning case, one energy level is 2 kWh and $E=28$, giving a total capacity of 56 kWh, which is within the range of compact electric and autonomous sweepers \citep{boschungUrbanSweeperS20,rascoLynxCharge2000,trombiaFree}. The on-site service stage corresponds to continuous cleaning over a local area. A one-level service takes 15 minutes, so the implied service-stage energy rate is 8 kW. This rate is close to the average operating power implied by the battery capacity and working time of compact electric sweepers, such as the RASCO LYNX Charge 2000, and is also consistent with the operating range of autonomous street sweepers \citep{rascoLynxCharge2000,trombiaFree}.

The return-to-station replenishment time $\tau^\mathrm{rc}$ is modeled as a fast operational turnaround rather than a full plug-in charging time. For the mobile-charging case, it represents battery-module exchange at the station and is set to 6 minutes. For the drone-inspection and autonomous-cleaning cases, it captures fast battery replacement, docking, and dispatch preparation, and is set to 4 and 10 minutes, respectively.

\begin{table}[htbp]
\centering
\caption{Parameter settings for the three real-case applications.}
\label{tab:real_case_settings}
\small
\begin{tabular}{lccc}
\toprule
Parameter & Mobile charger & Inspection drone & Autonomous cleaning robot \\
\midrule
Energy levels $E$ & 10 & 10 & 28 \\
Energy per level & 16 kWh & 10 Wh & 2 kWh \\
Battery capacity & 160 kWh & 100 Wh & 56 kWh \\
Travel speed & 16 km/hour & 36 km/hour & 5 km/hour \\
Service time per level & 20 min & 10 min & 15 min \\
Service-stage energy rate & 48 kW & 60 W & 8 kW \\
Station replenishment time & 6 min & 4 min & 10 min \\
\bottomrule
\end{tabular}
\end{table}

\section{Additional Experimental Results}\label{sec:app_exp_results}

\subsection{Additional Evidence on the Role of Energy}\label{sec:app_role_energy}

\begin{figure}[!ht]
    \centering
    \includegraphics[width=\textwidth]{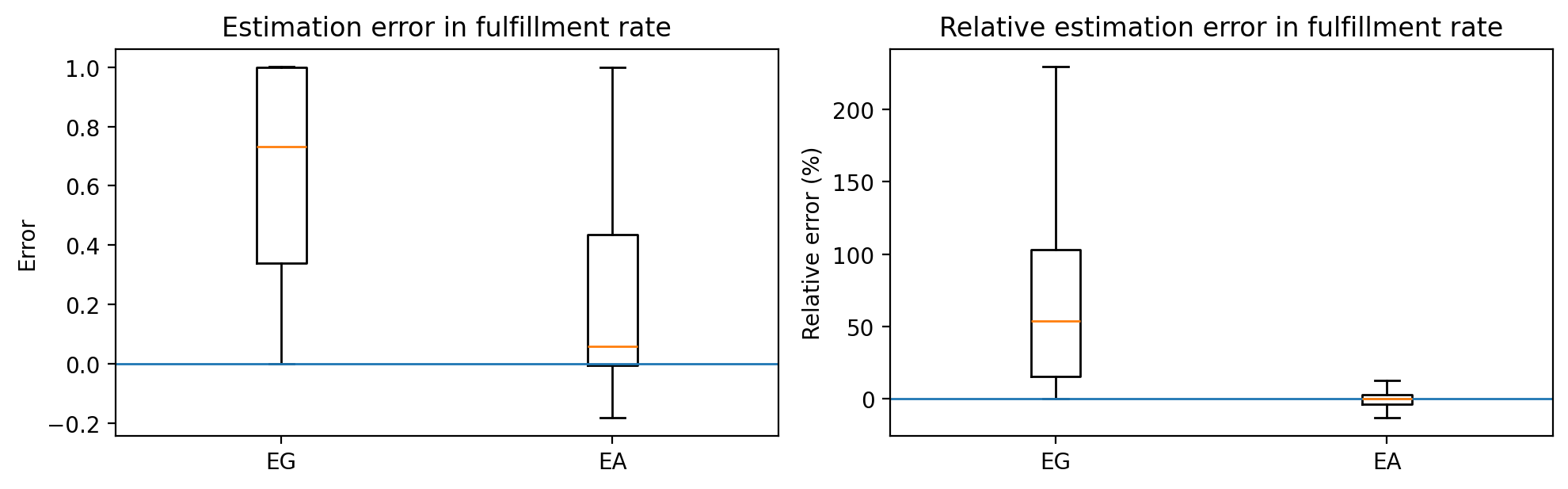}
    \caption{Distribution of estimation errors in fulfillment probability.}\label{fig:u_je_boxplots}
    \smallskip
    \noindent\begin{minipage}{\textwidth}
    \footnotesize
    \raggedright
    \emph{Notes.} The left panel reports the raw estimation error $u_{je}^{m}-u_{je}^{\mathrm{sim}}$, and the right panel reports the relative estimation error $100\times (u_{je}^{m}-u_{je}^{\mathrm{sim}})/u_{je}^{\mathrm{sim}}$ for the energy-agnostic (EG) and energy-aware (EA) models.
    \end{minipage}
\end{figure}

This subsection reports additional evidence on the estimation accuracy of the energy-aware (EA) and energy-agnostic (EG) models. Whereas Figure~\ref{fig:ffr_decomp} in the main text focuses on the positive part of the error that creates false promises, Figure~\ref{fig:u_je_boxplots} shows the full distribution of estimation errors in fulfillment probability across all $(j,e)$ cells.

Figure~\ref{fig:u_je_boxplots} shows that the benefit of EA is not limited to reducing the positive error mass that drives false fulfillment. EG overestimates $u_{je}$ in 96.1\% of all $(j,e)$ cells, indicating a strong upward bias in predicted service capability. By contrast, EA exhibits both positive and negative deviations (71.9\% positive and 28.1\% negative), and its relative error distribution is much more concentrated, with median $-0.38\%$ and interquartile range approximately $[-4.26\%,\,2.46\%]$. Thus, once energy states are modeled explicitly, the approximation becomes not only less optimistic but also substantially more accurate overall.

\subsection{Additional Evidence on the Mechanism of Zoning}\label{sec:app_zoning_mechanism}

This subsection reports supplementary evidence on the travel-side mechanism behind zoning. The main text shows that zoning hurts under light load but helps under heavy load. Table~\ref{tab:travel_per_served} isolates one key channel behind this reversal by reporting travel time per completed service under different load levels and zoning intensities.

\begin{table*}[t]
\centering
\footnotesize
\caption{Travel time per completed service, $\mathrm{TPS}(\alpha,A)$.}\label{tab:travel_per_served}
\setlength{\tabcolsep}{7pt}
\begin{tabular}{rcccccc}
\hline
$A$ & $\alpha=1$ & $\alpha=2$ & $\alpha=3$ & $\alpha=4$ & $\alpha=5$ & $\alpha=6$ \\
\hline
1 & 4.38 & 8.18 & 13.40 & 18.01 & 20.59 & 21.64 \\
2 & 4.97 & 8.73 & 11.72 & 14.27 & 15.38 & 16.50 \\
3 & 5.18 & 7.75 & 10.43 & 12.08 & 13.34 & 13.93 \\
4 & 4.39 & 6.64 & 8.08 & 9.07 & 9.38 & 9.67 \\
\hline
\end{tabular}

\vspace{4pt}
\noindent\begin{minipage}{\textwidth}
\footnotesize
\raggedright
\emph{Notes.} $\mathrm{TPS}(\alpha,A)=T_{\text{travel}}(\alpha,A)/N_{\text{served}}(\alpha,A)$, where $T_{\text{travel}}(\alpha,A)$ denotes total fleet travel time under configuration $(\alpha,A)$ and $N_{\text{served}}(\alpha,A)$ denotes the number of completed services. Higher values indicate that more travel effort is required per completed job.
\end{minipage}
\end{table*}

Table~\ref{tab:travel_per_served} shows that, under pooling ($A=1$), travel time per completed service rises sharply with load, from 4.38 at $\alpha=1$ to 21.64 at $\alpha=6$. Stronger zoning substantially mitigates this increase at moderate and high load: for example, at $\alpha=6$, $\mathrm{TPS}$ falls from 21.64 under pooling to 9.67 under $A=4$. By contrast, under light load the travel burden is already modest under pooling, so zoning has limited room to create meaningful travel savings. This is why the benefit of zoning emerges only when demand becomes sufficiently dense that travel containment dominates the loss of matching flexibility.

\subsection{Full Results for Joint Design of Battery Capacity and Zoning}\label{sec:app_joint_design}

This subsection reports the profit rate underlying the joint design discussion in Section~\ref{sec:joint_battery_zoning}. For each battery level $E$ and load level $\alpha$, Table~\ref{tab:toronto_joint_energy_zoning} reports the heuristic profit rate $R$ under $A=1,2,3,4$, where $A=1$ is the no-zoning benchmark, together with the adjacent profit rate increments
$
\Delta_{21}=R(A=2)-R(A=1),
\Delta_{32}=R(A=3)-R(A=2),
\Delta_{43}=R(A=4)-R(A=3).
$

\begin{table}[!ht]
\centering
\scriptsize
\caption{Heuristic revenue under joint battery and zoning design in the Toronto case.}\label{tab:toronto_joint_energy_zoning}
\begin{tabular}{ccrrrrrrr}
\toprule
$E$ & $\alpha$ & $R(A{=}1)$ & $R(A{=}2)$ & $R(A{=}3)$ & $R(A{=}4)$ & $\Delta_{21}$ & $\Delta_{32}$ & $\Delta_{43}$ \\
\midrule
10 & 1  & 2.1303 & 2.0924 & 2.0501 & 2.0179 & -0.0379 & -0.0422 & -0.0322 \\
10 & 4  & 4.4768 & 4.7462 & 4.8605 & 4.9251 &  0.2694 &  0.1143 &  0.0647 \\
10 & 7  & 4.2382 & 4.8935 & 5.2703 & 5.5445 &  0.6552 &  0.3768 &  0.2742 \\
10 & 10 & 4.1295 & 4.8939 & 5.3775 & 5.7475 &  0.7644 &  0.4836 &  0.3700 \\
\midrule
14 & 1  & 2.1241 & 2.0820 & 2.0361 & 2.0007 & -0.0421 & -0.0459 & -0.0354 \\
14 & 4  & 4.5787 & 4.8090 & 4.8964 & 4.9464 &  0.2303 &  0.0874 &  0.0500 \\
14 & 7  & 4.3640 & 5.0028 & 5.3577 & 5.6169 &  0.6388 &  0.3549 &  0.2593 \\
14 & 10 & 4.2552 & 5.0150 & 5.4836 & 5.8426 &  0.7598 &  0.4686 &  0.3590 \\
\midrule
18 & 1  & 2.1199 & 2.0749 & 2.0267 & 1.9895 & -0.0449 & -0.0482 & -0.0372 \\
18 & 4  & 4.6281 & 4.8377 & 4.9114 & 4.9541 &  0.2096 &  0.0737 &  0.0426 \\
18 & 7  & 4.4291 & 5.0584 & 5.4023 & 5.6529 &  0.6293 &  0.3439 &  0.2506 \\
18 & 10 & 4.3213 & 5.0779 & 5.5382 & 5.8915 &  0.7566 &  0.4603 &  0.3533 \\
\midrule
22 & 1  & 2.1174 & 2.0709 & 2.0214 & 1.9834 & -0.0465 & -0.0494 & -0.0380 \\
22 & 4  & 4.6591 & 4.8573 & 4.9237 & 4.9629 &  0.1981 &  0.0664 &  0.0392 \\
22 & 7  & 4.4707 & 5.0949 & 5.4329 & 5.6792 &  0.6242 &  0.3380 &  0.2463 \\
22 & 10 & 4.3643 & 5.1193 & 5.5751 & 5.9255 &  0.7550 &  0.4558 &  0.3504 \\
\bottomrule
\end{tabular}

\vspace{4pt}
\noindent\begin{minipage}{\textwidth}
\footnotesize
\raggedright
\emph{Notes.} For each $(E,\alpha,A)$, $R(A)$ (in CAD/min) is the profit rate produced by the heuristic algorithm. The adjacent increments $\Delta_{21}$, $\Delta_{32}$, and $\Delta_{43}$ report the marginal profit-rate gain from moving to the next tighter zoning level.
\end{minipage}
\end{table}

The full table confirms three patterns emphasized in the main text. First, the preferred zoning level is stable over the tested battery range: $A=1$ is best at $\alpha=1$, whereas $A=4$ is best at $\alpha=4,7,10$. Second, battery capacity and tighter zoning are partial substitutes, because the profit-rate gap between weaker and stronger zoning narrows as $E$ increases. Third, zoning has the larger effect on $R$: at moderate and high load, moving to a tighter zoning level typically produces a larger gain than increasing battery capacity within a fixed zoning design.

\putbib[reference]
\end{bibunit}

%
%
%

\end{document}